\documentclass[twocolumn,prd,a4paper,nobibnotes,nofootinbib,superscriptaddress,preprintnumbers]{revtex4}%
\usepackage[dvipdfmx]{graphicx,xcolor}
\usepackage[section]{placeins}
\usepackage{amsfonts,amsmath,amssymb}
\usepackage{array}
\usepackage{bm}
\usepackage{caption}
\usepackage{CJK}
\usepackage{float}
\usepackage{mathrsfs}
\usepackage{subcaption}
\DeclareMathOperator{\tr}{tr}
\DeclareMathOperator{\re}{Re}

\allowdisplaybreaks

\begin{document}
\preprint{CHIBA-EP-259}
\preprint{KEK Preprint 2023-27}
\title{Gauge-independent transition separating confinement and Higgs phases in lattice SU(2) gauge theory with a scalar field in the fundamental representation
}
\author{Ryu \surname{Ikeda}}
\email{cdna0955@chiba-u.jp}
\affiliation{Department of Physics, Graduate School of Science and Engineering, Chiba University, Chiba 263-8522, Japan}
\author{Seikou \surname{Kato}}
\email{skato@oyama-ct.ac.jp}
\affiliation{Oyama National College of Technology, Oyama 323-0806, Japan}
\author{Kei-Ichi \surname{Kondo}}
\email{kondok@faculty.chiba-u.jp}
\affiliation{Department of Physics, Graduate School of Science, Chiba University, Chiba 263-8522, Japan}
\author{Akihiro \surname{Shibata}}
\email{akihiro.shibata@kek.jp}
\affiliation{Computing Research Center, High Energy Accelerator Research Organization (KEK), Tsukuba 305-0801, Japan}
\affiliation{Department of Accelerator Science, SOKENDAI (The Graduate University for Advanced Studies), Tsukuba 305-0801, Japan}
\keywords{gauge-scalar model, gauge-independent BEH mechanism, confinement }
\pacs{PACS number}

\begin{abstract}

According to the preceding studies, the lattice SU(2) gauge-scalar model with a single scalar field in the fundamental representation of the gauge group has a single confinement-Higgs phase where confinement and Higgs regions are subregions of an analytically continued single phase and there are no thermodynamic phase transitions, which is a well-known consequence of the Osterwalder-Seiler-Fradkin-Shenker theorem.  
In this paper, we show that we can define new types of gauge-invariant operators by combining the original fundamental scalar field and the so-called color-direction field which is obtained by change of field variables based on the gauge-covariant decomposition of the gauge field due to Cho-Duan-Ge-Shabanov and Faddeev-Niemi.
By performing the numerical simulations on the lattice without any gauge fixing, we reproduce the conventional thermodynamic transition line in the weak gauge coupling, and moreover we find a new transition line detected by the new gauge-invariant operators which separates the confinement-Higgs phase into two parts, confinement phase and the Higgs phase, in the strong gauge coupling.
All results are obtained in the gauge-independent way, since no gauge fixing has been imposed in the numerical simulations.  
Moreover, we discuss a physical interpretation for the new transition from the viewpoint of the realization of a global symmetry.
\end{abstract}
\maketitle

\section{Introduction}

It is well known that the lattice SU(2) gauge-scalar model with a single scalar field in the fundamental representation of the gauge group has a single confinement-Higgs phase, according to the Osterwalder-Seiler-Fradkin-Shenker (OSFS) theorem \cite{OsterwalderSeiler78,FradkinShenker79,BanksRabinovici79}: Confinement and Higgs regions are subregions of an analytically continued single phase and there are no thermodynamic phase transitions between them.   

However, physics to be realized in the confinement region and the Higgs region are quite different. 
Therefore, there have been some attempts to elucidate the distinction between the two regions despite the absence of thermodynamic transition. See e.g., \cite{Maas19} for a review.
Recently, Greensite and Matsuyama \cite{Greensite-Matsuyama18,Greensite-Matsuyama20} proposed that the two regions can be discriminated by the symmetric or broken realization of a global symmetry called the \textit{custodial symmetry} which is a global symmetry acting on the scalar field alone: 
In the Higgs phase the custodial symmetry is spontaneously broken, while in the confinement phase the custodial symmetry is unbroken.
Their proposal is quite interesting.  Therefore, it must be confirmed by the independent research.


In this paper, we propose new types of gauge-invariant composite operators which enable to discriminate between the confinement phase and the Higgs phase in the lattice SU(2) gauge-scalar model with the scalar field in the \textit{fundamental} representation. 
The new gauge-invariant composite operators are constructed through the gauge-independent procedure by combining the original fundamental scalar field and the \textit{color-direction field} which is obtained by change of field variables based on the gauge-covariant decomposition of the gauge field due to Cho-Duan-Ge-Shabanov \cite{Cho8081,DuanGe79,Shabanov99} and Faddeev-Niemi  \cite{FaddeevNiemi9907} (CDGSFN decomposition), see \cite{KKSS15} for a review. 
This type of operator was already introduced for investigating the phase structure of the lattice SU(2) gauge-scalar model with the scalar field in the \textit{adjoint} representation to show the existence of the transition line which divides the confinement phase into two parts \cite{ShibataKondo23} in addition to the transition line separating the Higgs phase from the confinement phase.

In order to give a complete understanding on the phase separation, we must show how to characterize the Higgs phase in the gauge-invariant manner.
Here we recall the conventional understanding of the Brout-Englert-Higgs (BEH) mechanism \cite{Higgs64,EnglertBrout64}:
The original continuous gauge symmetry is spontaneously broken as a consequence of nonvanishing vacuum expectation value of the scalar field. 
In the case of the fundamental scalar field, the original SU(2) gauge symmetry is completely broken spontaneously if the fundamental scalar field acquires a nonvanishing vacuum expectation value. 
Then three massless Nambu-Goldstone particles must appear according to the Nambu-Goldstone theorem.  Nevertheless, they are absorbed into the original massless gauge bosons to make them massive and consequently no massless Nambu-Goldstone particles appear in the spectrum.

However, the conventional understanding of the BEH mechanism has some difficulties. 
One is that the vacuum expectation value of the fundamental scalar field vanishes unless the gauge-fixing condition is imposed, because the fundamental scalar field is not gauge invariant. 
This is a consequence of the Elitzur theorem \cite{Elitzur75}: the vacuum expectation value of the gauge noninvariant operator vanishes and the gauge symmetry cannot be spontaneously broken without gauge-fixing.  
In addition to this issue, whether or not the vacuum expectation value of the scalar field vanishes depend on the gauge choice even after the gauge-fixing condition is imposed.
Thus the conventional understanding of the BEH mechanism is not given in the gauge-invariant way.

To avoid these issues, one of the authors has proposed the \textit{gauge-independent description of the BEH mechanism} in the gauge-invariant manner (for the fundamental scalar field \cite{Kondo18} and the adjoint scalar field \cite{Kondo16}) which works irrespective of the gauge choice by introducing neither the spontaneous gauge symmetry breaking nor nonvanishing vacuum expectation value of the scalar field.
In this way the Higgs phase can be characterized in the gauge-independent way without any gauge-fixing. 
Under the new understanding of the BEH mechanism, we can investigate both the confinement mechanism and the BEH mechanism in the gauge-fundamental scalar model in the gauge-independent manner.

Based on these observations, we investigate the phase structure for the above model by performing the numerical simulations on the lattice without any gauge fixing.
Then (i) we gauge-independently reproduce the conventional thermodynamic transition line in the weak gauge coupling region\cite{OsterwalderSeiler78,FradkinShenker79,LangRebbiVirasoro81,Bonati}, 
and (ii) we find a new transition line which separates the confinement-Higgs phase into two different phases, the confinement phase and the Higgs phase, in the strong gauge coupling region.
All of these results are obtained in the gauge-independent way, since no gauge fixing has been imposed in these numerical simulations to measure gauge-invariant operators.

Moreover, we examine whether or not the two regions are discriminated by the symmetric or broken realization of a global symmetry as suggested by \cite{Greensite-Matsuyama18,Greensite-Matsuyama20}: 
The global symmetry is spontaneously broken in the Higgs phase, while it is unbroken in the confinement phase.

At first glance the new transition line we found seems to contradict with the OSFS theorem. As discussed in detail in the text, indeed, our results are compatible with the theorem.
The existence of the transition line separating the confinement and Higgs phases in the gauge-matter model with the matter field in the fundamental representation will shed new light on the QCD phase diagram, e.g., quark-hadron continuity \cite{SW99}, see also \cite{Cherman20}. 
Notice that the SU(2) gauge-scalar model with the fundamental scalar has no 1-form symmetry \cite{Gaiotto}, in contrast to the adjoint scalar model.  Therefore, the spontaneous breaking of the 1-form symmetry cannot be used to distinguish the phases in this model. 

This paper is organized as follows. 
In Sec.II, we define the lattice SU(2) gauge-scalar model with a radially-fixed scalar field in the fundamental representation.
By introducing the lattice version \cite{Kondo08,Exactdecomp09} of the CDGSFN decomposition for the gauge variable, we define the color-direction field to construct new gauge-invariant operators to be measured. 
In Sec.III, we explain our method of numerical simulations and give the results of numerical simulations. 
By measuring the gauge-invariant composite operator composed of the fundamental scalar field and the color-direction field, we finally find that the confinement-Higgs phase is separated into two regions, the confinement phase and the Higgs phase. 
In Sec.IV, we discuss possible physical interpretation of the simulation results. 
The last section is devoted to conclusion and discussion.

In Appendix A, we give another formulation which enables us to give a gauge-independent description of the BEH mechanism for the SU(2) gauge-scalar model with a radially-fixed scalar field in the fundamental representation. 
In Appendix B, we give some details on technical points which guarantee that the order parameter we proposed really takes the zero or nonzero value across the new transition.

\section{SU(2) lattice gauge-scalar model with a scalar field in the fundamental representation}

\subsection{Lattice gauge-scalar action and global symmetry}

We introduce the lattice SU(2) gauge-scalar model with a single scalar field in the fundamental representation of the gauge group where the radial degrees of freedom of the scalar field is fixed (no Higgs modes). The action of this model with the gauge coupling constant $\beta$ and the scalar coupling constant $\gamma$ is given in the standard way by
\begin{align}
 &S[U,\hat{\Theta}]
  = S_G [U] + S_H [U,\hat{\Theta}] \, , \nonumber\\
 &S_G [U] 
  = \frac{\beta}{2} \sum_{x,\mu>\nu} \re \tr \left( \mathbf{1} - U_{x,\mu} U_{x+\mu,\nu} U_{x+\nu,\mu}^{\dagger} U_{x,\nu}^{\dagger} \right) , \nonumber\\
 &S_H [U,\hat{\Theta}] 
  = \frac{\gamma}{2} \sum_{x,\mu} \re \tr \left( \mathbf{1} - {\hat{\Theta}}_x^{\dagger} U_{x,\mu} {\hat{\Theta}}_{x+\mu} \right) ,
\label{action1}
\end{align}
where $U_{x,\mu} \in \mathrm{SU(2)}$ is a (group-valued) gauge variable on a link $\langle x,\mu \rangle$, and ${\hat{\Theta}}_x \in \mathrm{SU(2)}$ is a (matrix-valued) scalar variable in the fundamental representation on a site $x$ which obeys the unit-length (or radially fixed) condition: 
\begin{align}
{\hat{\Theta}}^{\dagger}_x {\hat{\Theta}}_x = \bm{1} ={\hat{\Theta}}_x {\hat{\Theta}}^{\dagger}_x  \, .
\end{align}

This action is invariant under the local $\mathrm{SU(2)}_\mathrm{local}$ gauge transformation and the global $\mathrm{SU(2)}_\mathrm{global}$ transformation for the link variable $U_{x,\mu}$ and the site variable ${\hat{\Theta}}_x$:
\begin{align}
  U_{x,\mu} &\mapsto U_{x,\mu}^{\prime} = {\Omega}_x U_{x,\mu} {\Omega}_{x+\mu}^{\dagger} \, , \quad {\Omega}_x \in \mathrm{SU(2)}_\mathrm{local} , \notag\\
  \hat{\Theta}_x &\mapsto \hat{\Theta}_x^{\prime} = {\Omega}_x \hat{\Theta}_x \Gamma, \, \quad \Gamma \in \mathrm{SU(2)}_\mathrm{global} .
\end{align}
The expectation value of an operator $\mathscr{O}$ in this model is defined by
\begin{align}
\langle \mathscr{O}[U,\hat{\Theta}]  \rangle = \frac{1}{Z} \int \mathcal{D} U \mathcal{D} \hat{\Theta} e^{-S[U,\hat{\Theta}]} \mathscr{O}[U,\hat{\Theta}] ,
\label{exvev1}
\end{align}
where the integration measures $\mathcal{D}U=\prod_{x,\mu} dU_{x,\mu}$ and $\mathcal{D} \hat{\Theta}=\prod_x d\hat{\Theta}_x$ are the invariant Haar measures for the SU(2) group and the normalization $\langle 1 \rangle =1$ is guaranteed by introducing the partition function $Z$. 
Therefore, this model has the $\mathrm{SU(2)}_\mathrm{local} \times \mathrm{SU(2)}_\mathrm{global}$ symmetry.
Notice that the global symmetry $\mathrm{SU(2)}_\mathrm{global}$ is acting on the scalar field alone.

In the na\"{i}ve continuum limit, the action (\ref{action1}) reduces to the continuum gauge-scalar model with a scalar field in the fundamental representation with a gauge coupling constant $g$ and a fixed-length condition ${\Theta}^{\dagger}_x {\Theta}_x = {\Theta}_x {\Theta}^{\dagger}_x =v^2\bm{1}$, where $\beta := 4/g^2$ and $\gamma := v^2/2$.

\subsection{Color-direction field, the reduction condition and the field decomposition }

In our investigations, the color-direction field defined shortly plays the key role. 
This new field was introduced in the framework of change of field variables \cite{KKSS15} which is originally based on the gauge-covariant decomposition of the gauge field due to Cho-Duan-Ge-Shabanov\cite{Cho8081,DuanGe79,Shabanov99} and Faddeev-Niemi\cite{FaddeevNiemi9907}. 
In what follows we give a very short review on this framework to see the origin of the color-direction field and its role played in understanding confinement, see \cite{KKSS15} for a thorough review. 

The \textit{color-direction field} on the lattice is a (Lie-algebra valued) site variable:
\begin{align}
 \bm{n}_x := n_x^A {\sigma}^A \in \mathrm{su(2)-u(1)} \quad (A=1,2,3) \, ,
\end{align}
where ${\sigma}^A$ are the Pauli matrices. $\bm{n}_x$ has the unit length $\bm{n}_x \cdot \bm{n}_x = 1$.
We require the transformation property of the color-direction field $\bm{n}_x$ as
\begin{align}
 \bm{n}_x \mapsto \bm{n}_x^{\prime} = {\Omega}_x \bm{n}_x {\Omega}_x^{\dagger} \, .
\end{align}

For a given gauge field configuration $\{ U_{x,\mu} \}$, we determine the color-direction field configuration $\{ \bm{n}_x \}$ (as the unique configuration up to the global color rotation) by minimizing the so-called \textit{reduction functional} $F_\mathrm{red} [\bm{n} ; U]$ under the gauge transformations:
\begin{align}
 F_\mathrm{red} [\{ \bm{n}\}  ; \{ U\}]
   &:= \sum_{x,\mu} \frac{1}{2} \tr \left[ {\left( D_{\mu} [U] \bm{n}_x \right)}^{\dagger} \left( D_{\mu} [U] \bm{n}_x \right) \right] \notag\\
   &= \sum_{x,\mu} \tr \left( \bm{1} - \bm{n}_x U_{x,\mu} \bm{n}_{x+\mu} U_{x,\mu}^{\dagger} \right) \, .
\label{red}
\end{align}
In this way, a set of color-direction field configurations $\{ \bm{n}_x \}$ is obtained as the (implicit) functional of the original link variables $\{ U_{x,\mu} \}$, which is written symbolically as 
\begin{align}
 \bm{n}^*= \underset{\bm{n}}{\text{argmin}} \, F_\mathrm{red} [\{ \bm{n}\}  ; \{ U\}].
\label{redc2}
\end{align}
This construction shows the nonlocal nature of the color-direction field. 

By introducing the color-direction field, we obtain the deformed theory in which the expectation value of an operator $\mathscr{O}$ including the color-direction field is calculated according to 
\begin{align}
& \langle \mathscr{O}[U,\hat{\Theta},\bm{n} ]  \rangle \nonumber\\
&= \frac{1}{Z} \int \mathcal{D} U \mathcal{D} \hat{\Theta} e^{-S[U,\hat{\Theta}]} \int \mathcal{D} \bm{n} \ \bm{\delta}(\bm{n} -\bm{n}^*)  \mathscr{O}[U,\hat{\Theta},\bm{n} ] ,
\label{exvev2}
\end{align}
where $\mathcal{D} \bm{n}=\prod_x d\bm{n}_x$ is the invariant measure for the color-direction field and $\bm{\delta}(\bm{n} -\bm{n}^*)$ is the Dirac delta function which plays the role of replacing $\bm{n}$ by $\bm{n}^*$ determined by (\ref{redc2}).
Notice that this definition reduces to the original one (\ref{exvev1}) if the operator $\mathscr{O}$ does not include the color-direction field $\mathscr{O}[U,\hat{\Theta}]$ because of $\int \mathcal{D} \bm{n} \bm{\delta}(\bm{n} -\bm{n}^*)=1$. 

We specify the symmetry anew in the deformed theory with nonlocality carried through the reduction procedure.
The deformed theory is invariant under the local $\mathrm{SU(2)}_\mathrm{local}$ gauge transformation and the global $\widetilde{\mathrm{SU(2)}}_\mathrm{global}$ transformation:
\begin{align}
  U_{x,\mu} &\mapsto U_{x,\mu}^{\prime} = {\Omega}_x U_{x,\mu} {\Omega}_{x+\mu}^{\dagger} \, , \quad {\Omega}_x \in \mathrm{SU(2)}_\mathrm{local} \,, \notag\\
  \hat{\Theta}_x &\mapsto \hat{\Theta}_x^{\prime} = {\Omega}_x \hat{\Theta}_x \Gamma, \, \quad \Gamma \in \widetilde{\mathrm{SU(2)}}_\mathrm{global} \,, \notag\\
 \bm{n}_x &\mapsto \bm{n}_x^{\prime} = {\Omega}_x \bm{n}_x {\Omega}_x^{\dagger} \, .
\label{utn}
\end{align}
Therefore, the deformed theory with the color-direction field has the $\mathrm{SU(2)}_\mathrm{local} \times \widetilde{\mathrm{SU(2)}}_\mathrm{global}$ symmetry. It should be noticed that the global symmetry $\widetilde{\mathrm{SU(2)}}_\mathrm{global}$ should be discriminated from the global symmetry $\mathrm{SU(2)}_\mathrm{global}$ of the original theory.

By way of the color-direction field, the original link variable $U_{x,\mu} \in \mathrm{SU(2)}$ is gauge-covariantly decomposable into the product of two field variables $X_{x,\mu}, V_{x,\mu} \in \mathrm{SU(2)}$:
\begin{align}
 U_{x,\mu} = X_{x,\mu} V_{x,\mu} \, .
\end{align}
For this purpose, we require that $V_{x,\mu}$ has the transformation law in the same form as the original link variable $U_{x,\mu}$:
\begin{align}
  V_{x,\mu} \mapsto V_{x,\mu}^{\prime} = {\Omega}_x V_{x,\mu} {\Omega}_{x+\mu}^{\dagger} \, ,
\end{align}
and that $X_{x,\mu}$ has the transformation law in the same form as the site variable $\bm{n}_x$:
\begin{align}
 X_{x,\mu} \mapsto X_{x,\mu}^{\prime} = {\Omega}_x X_{x,\mu} {\Omega}_x^{\dagger} \, .
\end{align}

This decomposition is uniquely determined by solving the \textit{defining equations} simultaneously (once the color-direction field is given):
\begin{align}
 &D_{\mu} [V] \bm{n}_x := V_{x,\mu} \bm{n}_{x+\mu} - \bm{n}_x V_{x,\mu} = 0 \, , \\
 &\tr \left( \bm{n}_x X_{x,\mu} \right) = 0 \, ,
\end{align}
where $D_{\mu} [V]$ denotes the covariant derivative in the adjoint representation.

Indeed, the exact solution is obtained in the following form:
\begin{align}
 V_{x,\mu} &= \tilde{V}_{x,\mu} / \sqrt{\frac{1}{2}\tr \left( \tilde{V}_{x,\mu}^{\dagger} \tilde{V}_{x,\mu} \right)} \notag\\
 \tilde{V}_{x,\mu} &:= U_{x,\mu} + \bm{n}_x U_{x,\mu} \bm{n}_{x+\mu} \, ,\\
 X_{x,\mu} &= U_{x,\mu} V_{x,\mu}^{\dagger} \, .
\end{align}
Therefore, all components $V_{x,\mu}, X_{x,\mu}$ of the decomposition are obtained for a given gauge field configuration $\{ U_{x,\mu} \}$ and the color-direction field configuration $\{ \bm{n}_x \}$ to be determined from $\{ U_{x,\mu} \}$. 
Finally, all components $\bm{n}_x, V_{x,\mu}, X_{x,\mu}$ of the decomposition are determined as the functional of $\{ U_{x,\mu} \}$ alone.

These new variables have been successfully used in understanding confinement based on the dual superconductor picture. For example, it has been shown in the pure gauge theory without the matter field that the restricted field $V$ gives the dominant part for quark confinement, while the remaining field $X$ corresponds to the massive modes and decouple in the low-energy region.  This gives the gauge-independent version of the Abelian dominance observed in the maximal Abelian gauge. See \cite{KKSS15} for more details and more applications of this reformulation of the gauge theory.

\subsection{Gauge-invariant operators}

We proceed to investigate the phase structure of the model using the framework explained in the above.
First, we measure the averages of the plaquette action density:
\begin{align}
 P
  = \frac{1}{6V} \sum_{x,\mu<\nu} \tr \left( U_{x,\mu} U_{x+\mu,\nu} U_{x+\nu,\mu}^{\dagger} U_{x,\nu}^{\dagger} \right) \, ,
\end{align}
and the scalar action density:
\begin{align}
 M
  = \frac{1}{4V} \sum_{x,\mu} \tr \left( \hat{\Theta}_x^{\dagger} U_{x,\mu} \hat{\Theta}_{x+\mu} \right) \, ,
\end{align}
where $V$ is the total number of sites on the lattice.

In addition to these averages, we measure the \textit{susceptibilities} of these quantities to detect the transition line more clearly:
\begin{align}
 {\chi}_P &= \langle P^2 \rangle - {\langle P \rangle}^2 \, , \\
 {\chi}_M &= \langle M^2 \rangle - {\langle M \rangle}^2 \, .
\end{align}
These quantities are measured in the original theory defined by (\ref{exvev1}).

Second, in the deformed theory defined by (\ref{exvev2}), it is possible to define a new gauge-invariant operator $\bm{r}_x$, which is constructed from the original fundamental scalar field $\hat{\Theta}_x$ and the color-direction field $\bm{n}_x$. First we introduce a local gauge-invariant  \textit{scalar-color composite field} $\bm{r}_x$:
\begin{align}
 \bm{r}_x &:= \hat{\Theta}_x^{\dagger} \bm{n}_x \hat{\Theta}_x = \bm{r}_x^\dagger \, , 
\end{align}
which however transforms under the global transformation in the covariant way :
\begin{align}
 \bm{r}_x &\mapsto \bm{r}_x^{\prime} = \Gamma^\dagger \bm{r}_x \Gamma \, .
\end{align}
Then we define the gauge-invariant \textit{scalar-color composite field density} $\bm{R}$ as the spacetime average of $\bm{r}_x$, which has the same global transformation property as $\bm{r}_x$:
\begin{align}
 \bm{R} &:= \frac{1}{V} \sum_x \bm{r}_x  = \frac{1}{V} \sum_x \hat{\Theta}_x^{\dagger} \bm{n}_x \hat{\Theta}_x = \bm{R}^\dagger \, , \\
 \bm{R} &\mapsto \bm{R}^{\prime} = \Gamma^\dagger \bm{R} \Gamma \, .
\end{align}
It should be remarked that $\bm{R}$ is not contained in the original action, in sharp contrast to the operators $P$ and $M$.%
\footnote{
Even if the operator to be measured is restricted to the original field variables, we can construct the other  operators which are gauge-invariant, but transform according to the adjoint representation of the global group $\Gamma \in \widetilde{\mathrm{SU(2)}}_\mathrm{global}$.
For example, for any positive integer $n=1,2, \cdots$, we have
\begin{align}
 \mathcal{O}^{P}_x := \hat{\Theta}_x^{\dagger} (U_{P_x})^n \hat{\Theta}_x \ \ \mapsto \ \Gamma^\dagger \mathcal{O}^{P}_x \Gamma \ , \notag 
\end{align}
where $U_{P_x}$ is the plaquette gauge variable starting at the site $x$ and ending at the same site $x$, e.g., $U_{P_x}=U_{x,\mu} U_{x+\mu,\nu} U_{x+\nu,\mu}^{\dagger} U_{x,\nu}^{\dagger}$.
In order to make the operator $\mathcal{O}^{P}_x$ invariant under the Lorentz transformation or the Euclidean rotation invariant, we must sum up all the possible directions of the plaquette $P_x$ specified by two indices $\mu$ and $\nu$. 
}

Notice that the matrix $\bm{R}$ is invariant under the gauge transformation, therefore, every component of the matrix $\bm{R}$ is gauge-invariant, but it is not invariant under the global transformation. 
Therefore, in order to show gauge-independently the spontaneous breaking of the global symmetry, we have only to measure one of the component of the matrix $\bm{R}$.
It is easy to show that $\bm{R}$ is written in the form of the Lie-algebra su(2) valued matrix:
\begin{align}
 \bm{R}
  := R^A {\sigma}^A
  = \begin{pmatrix} R^3 & R^1-i R^2 \\ R^1+i R^2 & -R^3 \end{pmatrix} \in \mathrm{su(2)} 
 \, .
\end{align}
Therefore, the $A$-component $R^A$ $(A=1,2,3)$ of $\bm{R}$ is obtained from $\bm{R}$ as
\begin{align}
 R^A = \frac{1}{2} \tr (\sigma^A \bm{R}) \, .  
\end{align}

In order to measure the average of the scalar-color composite field density $\bm{R}$, it is necessary to solve numerically the reduction condition (\ref{red}) to obtain the color-direction field configuration $\{ \bm{n}_x \}$. However, there are two types of ambiguity to determine uniquely the color-direction field configuration. 

(i) 
One ambiguity comes from the existence of the Gribov copies which give the local minimum of the reduction functional. 
In order to resolve this issue, namely, to avoid the local minima and to obtain the global minima, the reduction functional is minimized under the random initial configurations. 

(ii) 
The other ambiguity comes from the invariance of the reduction functional under the global sign flip $\{ \bm{n}_x \} \mapsto \{ -\bm{n}_x \}$. 
Consequently, the average of the operator including the color field can vanish due to cancellations between a configuration $\{ \bm{n}_x \}$ and the flipped one $\{ -\bm{n}_x \}$.  
To avoid this issue we propose to measure the average $\langle |R^A| \rangle$ using  the absolute value:
\begin{align}
 |R^A|
  &= \left|\frac{1}{2} \tr (\sigma^A \bm{R}) \right| = \left| \frac{1}{V} \sum_x  \frac{1}{2} \tr (\sigma^A R_x ) \right| \notag\\
  &= \left| \frac{1}{V} \sum_x \frac{1}{2} \tr (\sigma^A \hat{\Theta}_x^{\dagger} \bm{n}_x \hat{\Theta}_x ) \right| \, ,
\label{abr3}
\end{align}
which is invariant under the sign flip of $\{ \bm{n}_x \}$. 

In particular, $|R^3|$ is rewritten into the form:
\begin{align}
 |R^3| 
 &= \left| \frac{1}{V} \sum_x \frac12{\rm tr}(\bm{n}_x \hat{\Theta}_x \sigma^3 \hat{\Theta}_x^{\dagger}  ) \right| \notag\\
 &= \left| \frac{1}{V} \sum_x \frac12{\rm tr}(\bm{n}_x  \bm{\phi}_x  ) \right| , \ \ \bm{\phi}_x := \hat{\Theta}_x \sigma^3 \hat{\Theta}_x^{\dagger} \, .
\label{abr3a}
\end{align}
Here $\bm{\phi}_x$ can be identified with a composite adjoint scalar field constructed from the fundamental scalar field $\hat{\Theta}_x$. 
In fact, $\bm{\phi}_x$ transforms according to the adjoint representation under the gauge transformation: $\bm{\phi}_x := \hat{\Theta}_x \sigma^3 \hat{\Theta}_x^{\dagger} \mapsto {\Omega}_x \bm{\phi}_x {\Omega}_x^{\dagger}$. 
Consequently, $|R^3|$ has the same form as the gauge-invariant composite operator $|Q|$ introduced for investigating  the lattice SU(2) gauge-scalar model with the adjoint scalar field $\bm{\phi}$ to show the existence of the transition line which divides the confinement phase into two parts \cite{ShibataKondo23} in addition to the transition line separating the Higgs phase from confinement phase. 


The gauge-invariant quantity $R^A$ as a component of the gauge-invariant matrix $\bm{R}$ transforms under the global transformation as 
\begin{align}
 &R^A = \frac12 {\rm tr}(\sigma^A \bm{R}) \notag\\
 &\mapsto \ R^{A \prime} = \frac12 {\rm tr}(\sigma^A \bm{R}^{\prime}) \notag\\
 &\hspace{12.5mm} = \frac12 {\rm tr}(\sigma^A \Gamma^\dagger \bm{R} \Gamma) 
 = \frac12 {\rm tr}(\Gamma \sigma^A \Gamma^\dagger \bm{R} ) \, .
\end{align}
Therefore, $R^A$ is invariant $R^{A \prime} = R^A$ under the continuous subgroup $\widetilde{\mathrm{U(1)}}_{A \, \mathrm{global}}$ of the global group $\widetilde{\mathrm{SU(2)}}_\mathrm{global}$, because
\begin{align}
 &\Gamma \sigma^A \Gamma^\dagger = \sigma^A \notag\\
 &\Leftrightarrow \ \Gamma=\exp (i\theta_A \sigma^A) \in \widetilde{\mathrm{U(1)}}_{A \, \mathrm{global}} \subset \widetilde{\mathrm{SU(2)}}_\mathrm{global} \notag\\
 &\qquad \text{(no sum over $A$)} \, .
\end{align}
This means that, if $\langle |R^A| \rangle$ has a nonvanishing value, the global group $\widetilde{\mathrm{SU(2)}}_\mathrm{global}$ is spontaneously broken to $\widetilde{\mathrm{U(1)}}_{A \, \mathrm{global}}$. 
However, this partial breaking does not give the true spontaneous symmetry breaking, because this breaking depends on the specific choice $A$ in the Lie algebra and there are no common subgroups for all $\widetilde{\mathrm{U(1)}}_{A \, \mathrm{global}}$ for $A=1,2,3$.

Thus, we need to take into account all the components on equal footing simultaneously to examine the spontaneous breaking of the global symmetry correctly.
From this viewpoint, we define the na\"ive gauge-invariant \textit{norm} as an order parameter:
\begin{align}
 {\left\| \bm{R} \right\|}_n  := \left( \sum_{A=1}^{3}  |R^A|^n \right)^{1/n} \, ,
\end{align}
which is expected to reflect the correlation between the color-direction field $\bm{n}_x$ and the fundamental scalar field $\hat{\Theta}_x$, and detect the spontaneous breaking of the global symmetry $\widetilde{\mathrm{SU(2)}}_\mathrm{global}$.

The $n=1$ case is just the sum of all the components:
\begin{align}
 {\left\| \bm{R} \right\|}_1  = \sum_{A=1}^{3}  |R^A| \, .
\label{abr3d}
\end{align}
This operator is not invariant under any continuous subgroup of the global group $\widetilde{\mathrm{SU(2)}}_\mathrm{global}$, and hence can be used to show the complete spontaneous breaking of the global symmetry $\widetilde{\mathrm{SU(2)}}_\mathrm{global}$. 

The $n=2$ case is written as
\begin{align}
 {\left\| \bm{R} \right\|}_2
  &= \left( \sum_{A=1}^{3}  |R^A|^2 \right)^{1/2} \notag\\
  &= \sqrt{  {(R^1)}^2 + {(R^2)}^2 + {(R^3)}^2 } = \sqrt{\bm{R} \cdot \bm{R}} \, .
\label{abr3e}
\end{align}
This is equivalent to consider 
the scalar-color composite density \textit{norm} $\left\|  \bm{R} \right\|_2$ defined by
\begin{align}
  {\left\| \bm{R} \right\|}_2 &= \sqrt{\frac{1}{2} \tr ( \bm{R}^{\dagger} \bm{R} )} \, , \label{scal} 
\end{align}
which is invariant under both the local gauge and global transformations:
\begin{align}
 {\left\| \bm{R} \right\|}_2 &\mapsto \left\| \bm{R}^{\prime} \right\|_2 = {\left\| \bm{R} \right\|}_2 \, .
\end{align}
The ambiguity of the global sign flipping is automatically avoided for $\langle {\left\| \bm{R} \right\|}_2 \rangle$ by virtue of the invariance of ${\left\| \bm{R} \right\|}_2$ under the global sign flipping, due to the fact that ${\left\| \bm{R} \right\|}_2$ is quadratic in $\bm{n}_x$:
\begin{align}
 {\left\| \bm{R} \right\|}_2
  &= \bigg[ \frac12 \tr \bigg\{ {\bigg( \frac{1}{V} \sum_x \hat{\Theta}_x^{\dagger} \bm{n}_x \hat{\Theta}_x \bigg)}^{\dagger} \notag\\
  &\hspace{15mm} \times \bigg( \frac{1}{V} \sum_y \hat{\Theta}_y^{\dagger} \bm{n}_y \hat{\Theta}_y \bigg) \bigg\} \bigg]^{1/2} \, .
\end{align}
To see the meaning of ${\left\| \bm{R} \right\|}_2$, we obtain the eigenvalues of $\bm{R}$  by solving the characteristic equation for the eigenvalue problem:
\begin{align}
 0 &= \det (\bm{R} - {\lambda} \bm{1})
  = \begin{vmatrix} R^3-{\lambda}  & R^1-iR^2 \\ R^1+iR^2 & -R^3-{\lambda}  \end{vmatrix} \notag\\
  &= {\lambda} ^2 - \left[ {(R^1)}^2 + {(R^2)}^2 + {(R^3)}^2 \right]
  \notag\\
  &= (\lambda-\lambda_{+})(\lambda-\lambda_{-}), \notag\\
 {\lambda}_{\pm} 
  &= \pm \sqrt{\bm{R}^2} := \pm \sqrt{  {(R^1)}^2 + {(R^2)}^2 + {(R^3)}^2 } \, .
\label{eigenv}
\end{align}
Therefore, the scalar-color density $\bm{R}$ can be transformed into the diagonal form and the norm ${\left\| \bm{R} \right\|}_2 $ consists of two eigenvalues of the scalar-color density $\bm{R}$: $\lambda={\lambda}_{\pm} := \pm \sqrt{\bm{R}^2}$, which reproduces e.g., $\pm R^3$ by a global rotation.

Moreover, we define the \textit{susceptibilities} of the scalar-color average norm to detect the \textit{new} transition line:
\begin{align}
 {\chi}_{{\left\| \bm{R} \right\|}_n} := \langle {\left\| \bm{R} \right\|}_n^2 \rangle - \langle {\left\| \bm{R} \right\|}_n \rangle^2 \, .
\end{align}


\section{numerical simulation}

\subsection{Settings for the lattice simulation}

We performed the Monte Carlo simulations on the $8^4$ and $16^4$ lattice with the periodic boundary condition.
In each Monte Carlo sweep, the configuration of link variables $\{ U_{x,\mu} \}$ and site variables $\{ \hat{\Theta}_x \}$ were updated alternately by the pseudo heat bath method (with Kennedy-Pendleton method \cite{KennedyPendleton85} for large $\beta, \gamma$).
For a measurement with a set of couplings $(\beta,\gamma)$, we discarded first 5000 sweeps for thermalization and sampled configurations per 100 sweeps and stored 100 configurations.

For each link field configuration $\{ U_{x,\mu} \}$, we obtained numerically the resulting color-direction field configuration $\{ \bm{n}_x \}$ by using the iterative method with over-relaxation to solve the reduction condition.

The above simulations were performed for $9 \times 16 = 144$ sets of couplings $(\beta,\gamma)$.

\FloatBarrier

\subsection{Averages for $P$ and $M$}

For the purpose of scanning the transition line on the phase diagram, we performed the measurement for the expectation value $\langle \mathcal{O} \rangle$ of the gauge-invariants $\mathcal{O}$ on various $\beta = \mathrm{const.}$ and $\gamma = \mathrm{const.}$ lines. We identified the transition lines by detecting gaps or rises of the plots for $\langle \mathcal{O} \rangle$.

First we measured the plaquette action density $P$ and the scalar action density $M$ to determine the thermodynamic transition line which is expected to reproduce the well-known transition line originally found by numerical simulations in \cite{LangRebbiVirasoro81}.

\begin{figure*}[!t]
\centering
\begin{subfigure}{88mm}
  \centering\includegraphics[width=85mm]{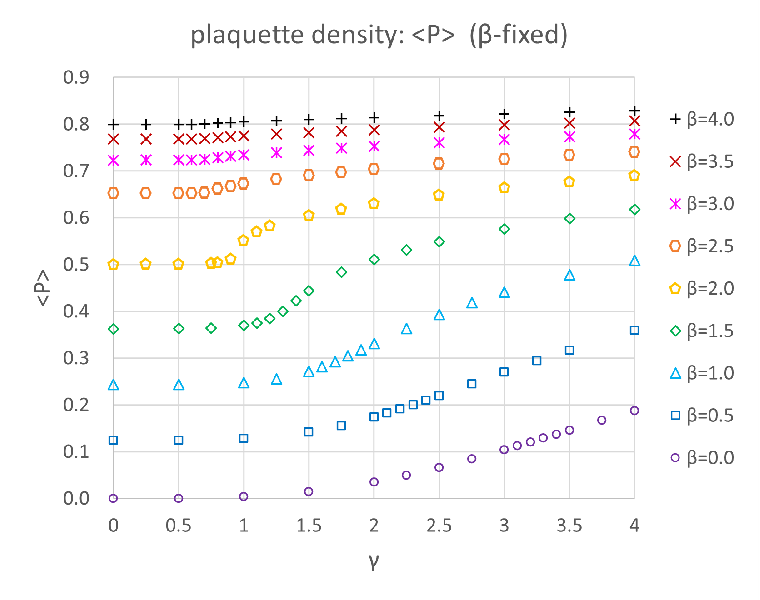}
  \label{pa1}
\end{subfigure}
\begin{subfigure}{88mm}
  \centering\includegraphics[width=85mm]{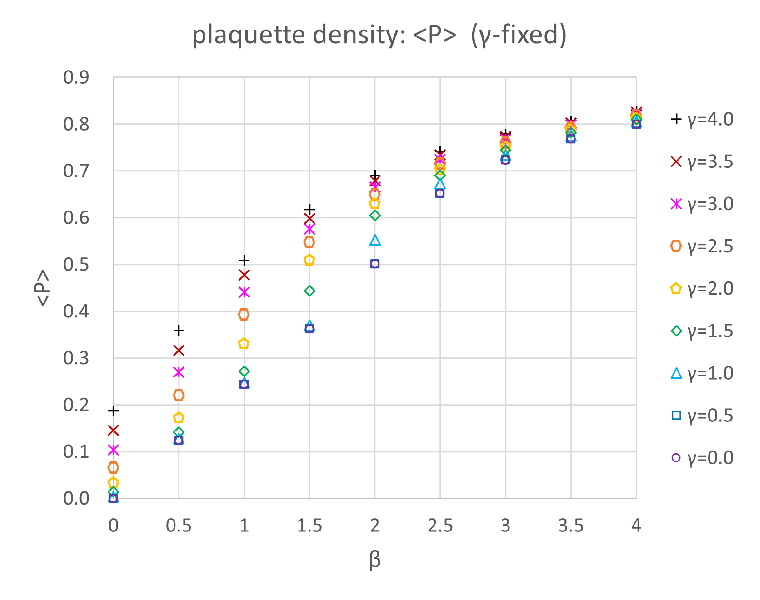}
  \label{pa2}
\end{subfigure}
\vspace{-3mm}\caption{Average of the plaquette action density $\langle P \rangle$ on the $16^4$ lattice: (left) $\langle P \rangle$ vs. $\gamma$ on various $\beta = \mathrm{const.}$ lines, (right) $\langle P \rangle$ vs. $\beta$ on various $\gamma = \mathrm{const.}$ lines.}
\label{pa12}
\end{figure*}

First, we determine the transition line from the plaquette action density $\langle P \rangle$. Figure \ref{pa12} shows the measurement results of $\langle P \rangle$ in the $\beta$-$\gamma$ phase plane. The left panel is the plots of $\langle P \rangle$ as functions of $\gamma$ on various $\beta = \mathrm{const.}$ lines, while the right panel is the plots of $\langle P \rangle$ as functions of $\beta$ on various $\gamma = \mathrm{const.}$ lines. In these plots, error bars are omitted because errors are too small to be indicated.

\begin{figure*}[!t]
\centering
\begin{subfigure}{88mm}
  \centering\includegraphics[width=85mm]{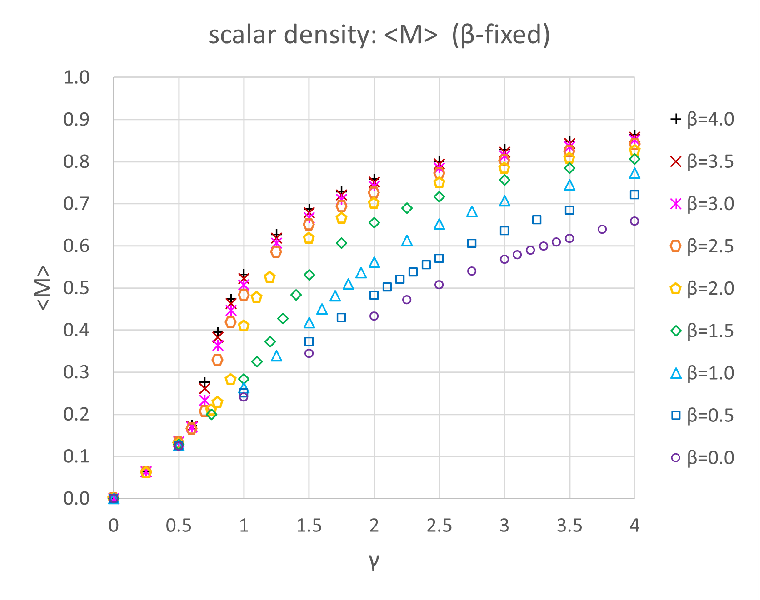}
  \label{sa1}
\end{subfigure}
\begin{subfigure}{88mm}
  \centering\includegraphics[width=85mm]{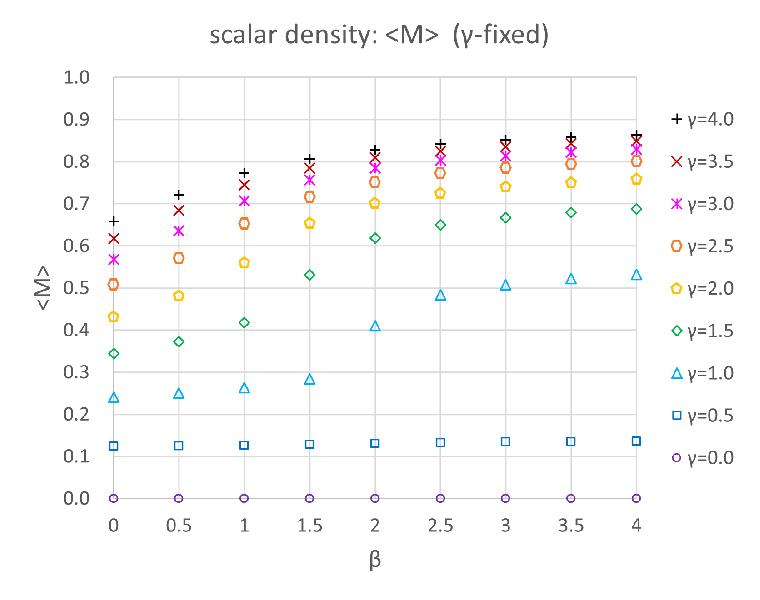}
  \label{sa2}
\end{subfigure}
\vspace{-3mm}\caption{Average of the scalar action density $\langle M \rangle$ on the $16^4$ lattice: (left) $\langle M \rangle$ vs. $\gamma$ on various $\beta = \mathrm{const.}$ lines, (right) $\langle M \rangle$ vs. $\beta$ on various $\gamma = \mathrm{const.}$ lines.}
\label{sa12}
\end{figure*}

Similarly, we determine the transition line from the scalar action density $\langle M \rangle$. Figure \ref{sa12} shows the measurement results of $\langle M \rangle$ in the $\beta$-$\gamma$ phase plane. The left panel is the plots of $\langle M \rangle$ as functions of $\gamma$ on various $\beta = \mathrm{const.}$ lines, while the right panel is the plots of $\langle M \rangle$ as functions of $\beta$ on various $\gamma = \mathrm{const.}$ lines.

By observing gaps in these plots, we obtained the transition lines in the left panel of Fig.\ref{psad} determined from the plaquette action density $\langle P \rangle$, and that in the right panel of Figure \ref{psad} from the scalar action density $\langle M \rangle$. The neighboring two observed points represent the pieces of the transition lines, and the error bars with the observed points were determined due to the interval of the simulation points.
Notice that these transition lines obtained from $\langle P \rangle$ and $\langle M \rangle$ agree with each other within the errors. Then we can conclude that we reproduced gauge-independently the transition line which was obtained in the specific gauge by the preceding studies \cite{LangRebbiVirasoro81,Bonati}, by performing the gauge-independent numerical simulations.

\subsection{Susceptibilities for $P$ and $M$}

We performed more measurements of gauge-invariants: the \textit{susceptibilities} (specific heat) ${\chi}_P$ of the plaquette action density $P$ and ${\chi}_M$ of the scalar action density $M$. We identified the transition lines by detecting peaks of the plots for these susceptibilities.

\begin{figure*}[!t]
\centering
\begin{subfigure}{88mm}
  \centering\includegraphics[width=85mm]{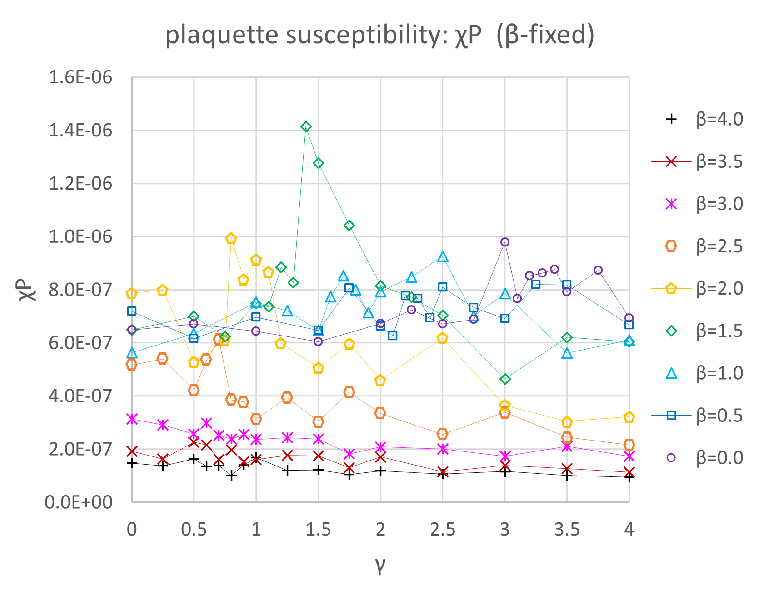}
  \label{ps1}
\end{subfigure}
\begin{subfigure}{88mm}
  \centering\includegraphics[width=85mm]{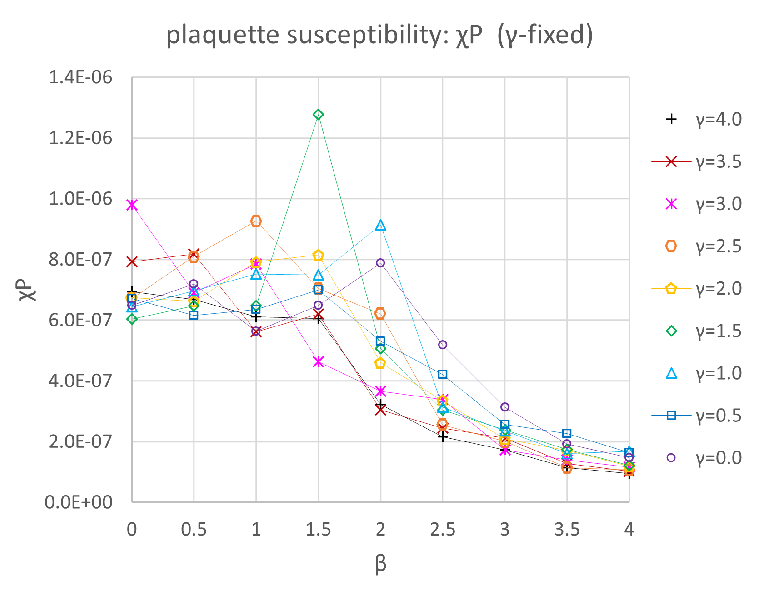}
  \label{ps2}
\end{subfigure}
\vspace{-3mm}\caption{Susceptibility ${\chi}_P$ of the plaquette action density $P$ on the $16^4$ lattice: (left) ${\chi}_P$ vs. $\gamma$ on various $\beta = \mathrm{const.}$ lines, (right) ${\chi}_P$ vs. $\beta$ on various $\gamma = \mathrm{const.}$ lines.}
\label{ps12}
\end{figure*}

We determine the transition line from the plaquette susceptibility ${\chi}_P := \langle P^2 \rangle - {\langle P \rangle}^2$ and the scalar susceptibility ${\chi}_M := \langle M^2 \rangle - {\langle M \rangle}^2$. Figure \ref{ps12} shows the measurement results of ${\chi}_P$ in the $\beta$-$\gamma$ phase plane. The left panel is the plots of ${\chi}_P$ as functions of $\gamma$ on various $\beta = \mathrm{const.}$ lines, while the right panel is the plots of ${\chi}_P$ as functions of $\beta$ on various $\gamma = \mathrm{const.}$ lines.

\begin{figure*}[!t]
\centering
\begin{subfigure}{88mm}
  \centering\includegraphics[width=85mm]{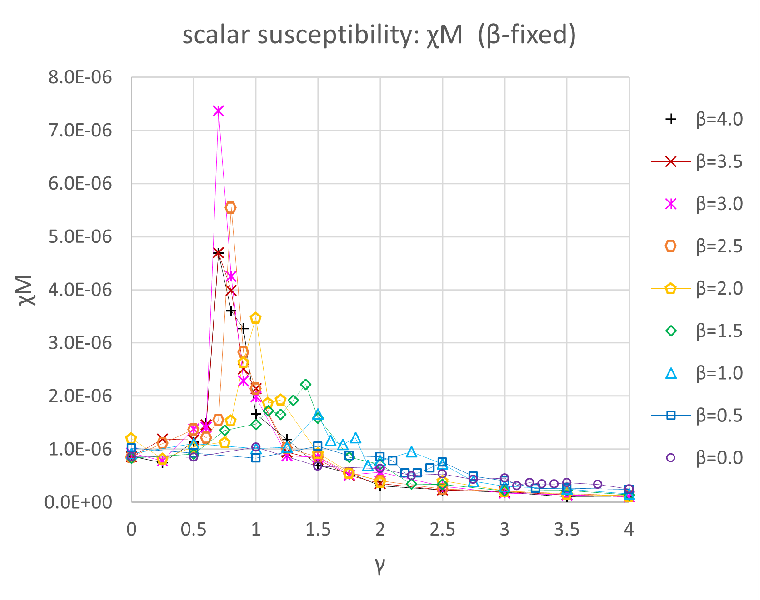}
  \label{ss1}
\end{subfigure}
\begin{subfigure}{88mm}
  \centering\includegraphics[width=85mm]{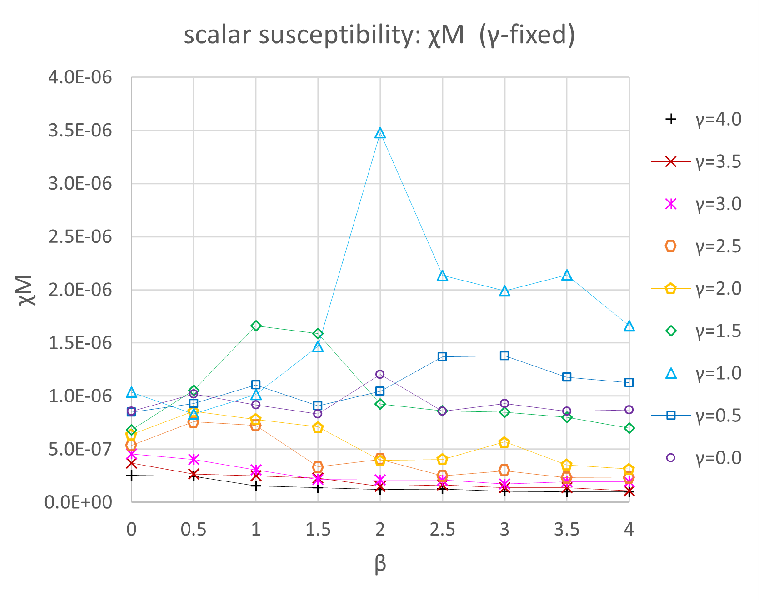}
  \label{ss2}
\end{subfigure}
\vspace{-3mm}\caption{Susceptibility ${\chi}_M$ of the scalar action density $M$ on the $16^4$ lattice: (left) ${\chi}_M$ vs. $\gamma$ on various $\beta = \mathrm{const.}$ lines, (right) ${\chi}_M$ vs. $\beta$ on various $\gamma = \mathrm{const.}$ lines.}
\label{ss12}
\end{figure*}

Moreover, Figure \ref{ss12} shows the measurement results of ${\chi}_M$ in the $\beta$-$\gamma$ phase plane. The left panel is the plots of ${\chi}_M$ as functions of $\gamma$ on various $\beta = \mathrm{const.}$ lines, while the right panel is the plots of ${\chi}_M$, as functions of $\beta$ on various $\gamma = \mathrm{const.}$ lines.

\begin{figure*}[!t]
\centering
\begin{subfigure}{88mm}
  \centering\includegraphics[width=84mm]{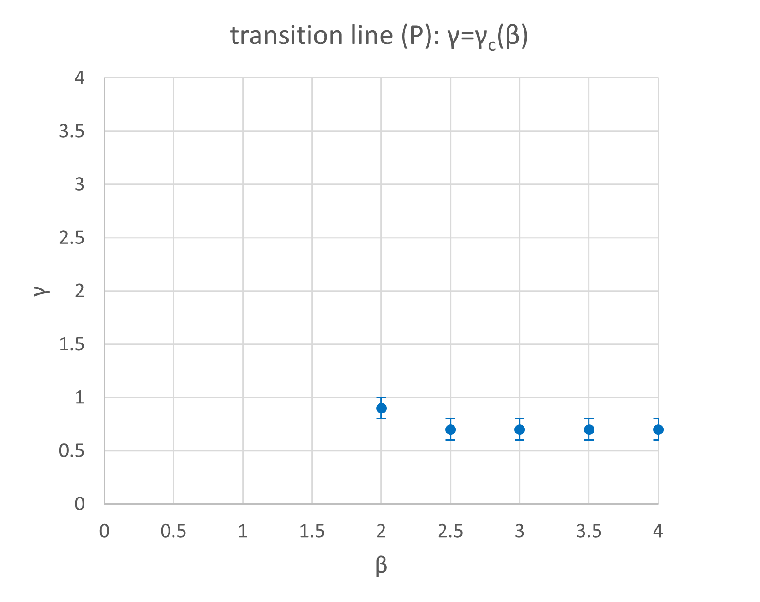}
  \label{pad}
\end{subfigure}
\begin{subfigure}{88mm}
  \centering\includegraphics[width=84mm]{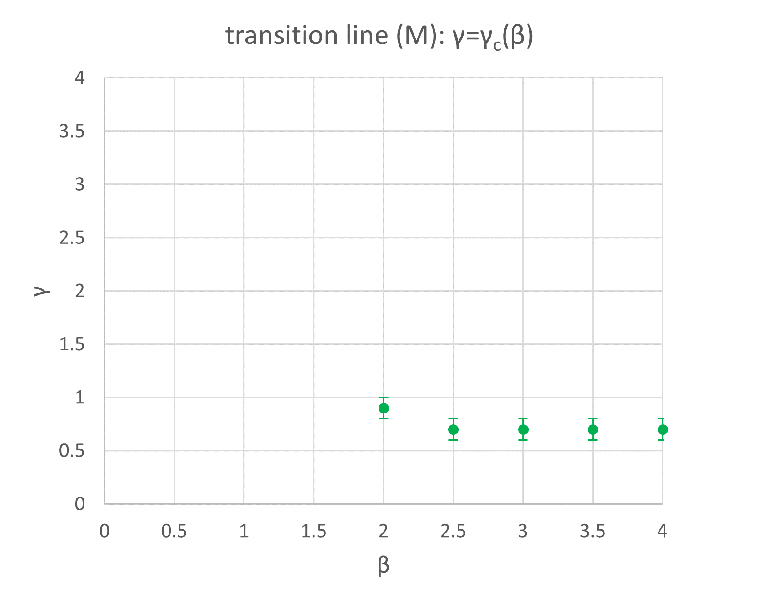}
  \label{sad}
\end{subfigure}
\vspace{-3mm}\caption{Transition lines $\gamma = {\gamma}_c (\beta)$ determined by the action densities on the $16^4$ lattice:  (left) $P$, (right) $M$.}
\label{psad}
\end{figure*}

\begin{figure*}[!t]
\centering
\begin{subfigure}{88mm}
  \centering\includegraphics[width=84mm]{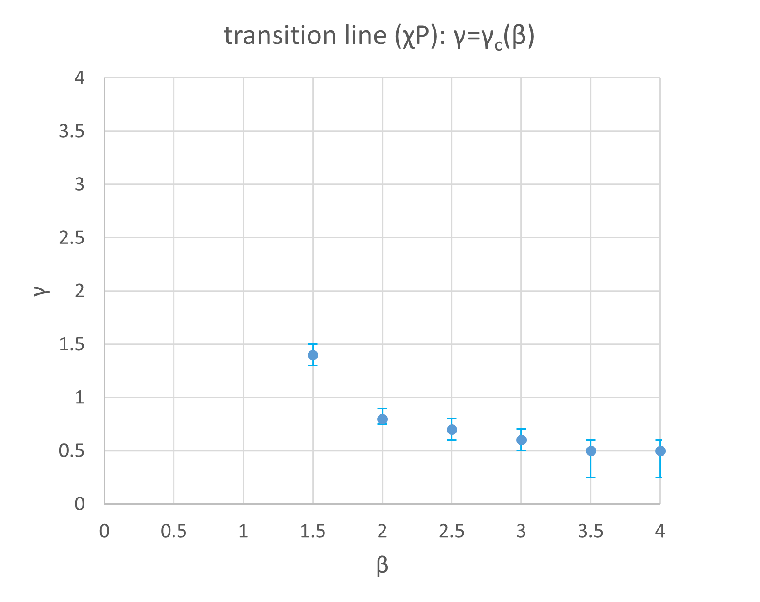}
  \label{psd}
\end{subfigure}
\begin{subfigure}{88mm}
  \centering\includegraphics[width=84mm]{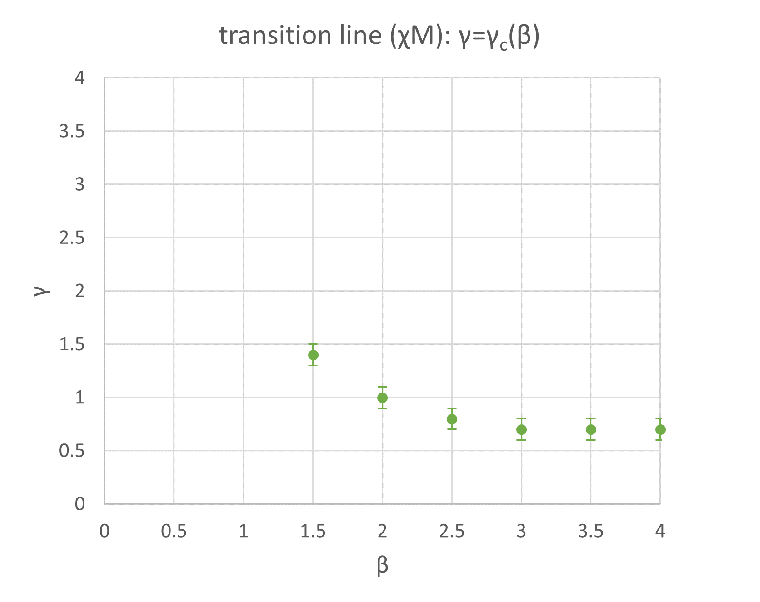}
  \label{ssd}
\end{subfigure}
\vspace{-3mm}\caption{Transition lines $\gamma = {\gamma}_c (\beta)$ determined by the susceptibilities on the $16^4$ lattice: (left) ${\chi}_P$ (right) ${\chi}_M$}
\label{pssd}
\end{figure*}

Figure \ref{pssd} shows the transition lines obtained by observing the peaks in these plots.
The transition line in the left panel of Fig.\ref{pssd} is determined from the plaquette susceptibility ${\chi}_P$ and that in the right panel of Fig.\ref{pssd} from the scalar susceptibility ${\chi}_M$. 
Notice that both transition lines obtained from ${\chi}_P$ and ${\chi}_M$ are consistent each other and coincide with the transition lines determined from $\langle P \rangle$ and $\langle M \rangle$ given in Fig.\ref{psad} within the errors.

These results reproduce the results obtained in the preceding studies \cite{LangRebbiVirasoro81,Bonati}.

\FloatBarrier

\begin{figure*}[!p]
\centering
\begin{subfigure}{88mm}
  \centering\includegraphics[width=85mm]{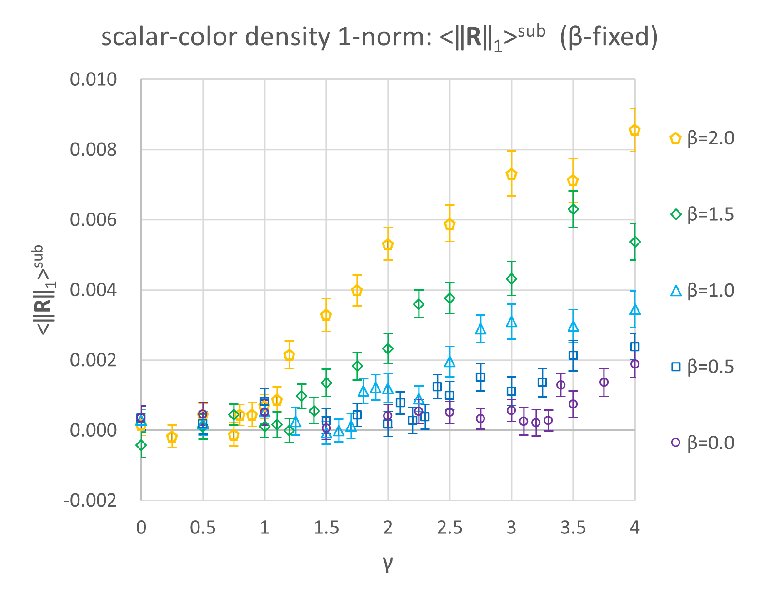}
  \label{ta1a}
\end{subfigure}
\begin{subfigure}{88mm}
  \centering\includegraphics[width=85mm]{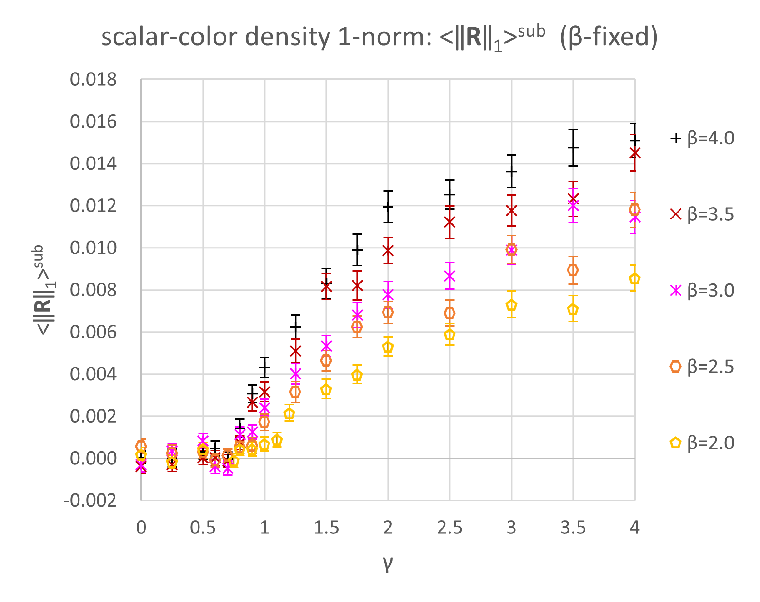}
  \label{ta1b}
\end{subfigure}\\
\begin{subfigure}{88mm}
  \centering\includegraphics[width=85mm]{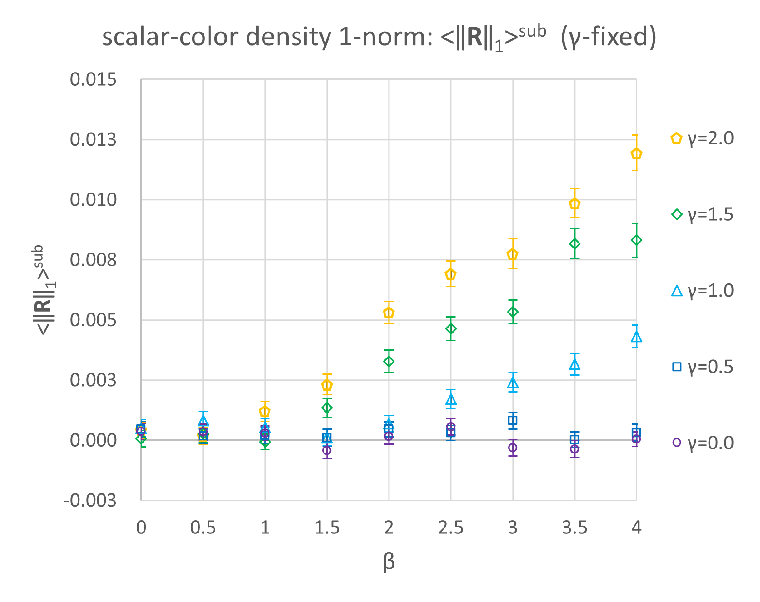}
  \label{ta2a}
\end{subfigure}
\begin{subfigure}{88mm}
  \centering\includegraphics[width=85mm]{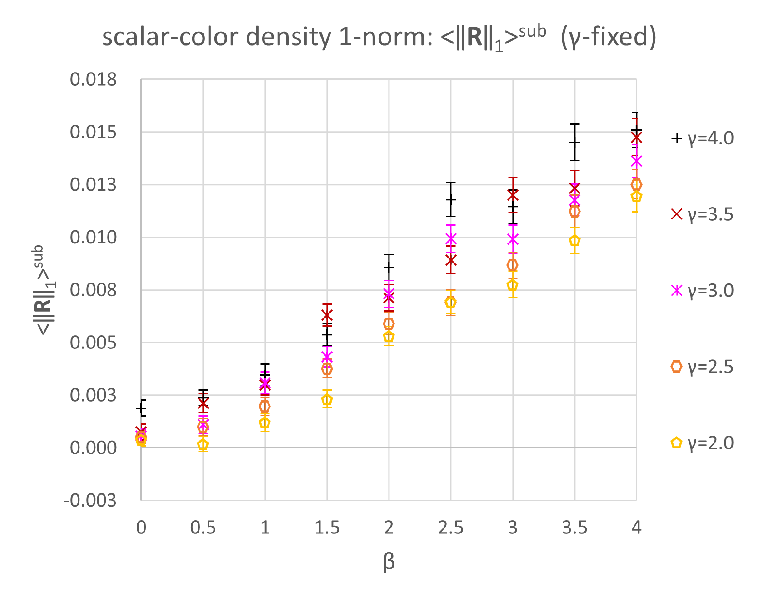}
  \label{ta2b}
\end{subfigure}
\caption{Average of the 1-norm of the scalar-color composite field density $\langle {\left\| \bm{R} \right\|}_1 \rangle^\mathrm{sub}$ on the $16^4$ lattice: (upper) $\langle {\left\| \bm{R} \right\|}_1 \rangle^\mathrm{sub}$ vs. $\gamma$ on various $\beta = \mathrm{const.}$ lines, (lower) $\langle {\left\| \bm{R} \right\|}_1 \rangle^\mathrm{sub}$ vs. $\beta$ on various $\gamma = \mathrm{const.}$ lines.} 
\label{ta12}
\end{figure*}

\begin{figure*}[!p]
\centering
\begin{subfigure}{88mm}
  \centering\includegraphics[width=84mm]{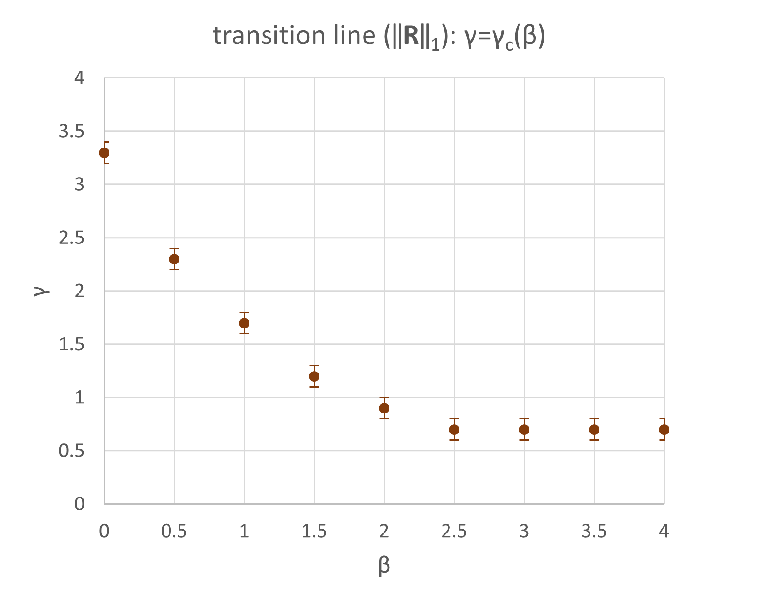}
\end{subfigure}
\caption{Transition lines $\gamma = {\gamma}_c (\beta)$ determined by the 1-norm of the scalar-color composite field density ${\left\| \bm{R} \right\|}_1$ on the $16^4$ lattice.}
\label{tad}
\end{figure*}

\subsection{Average for ${\left\| \bm{R} \right\|}_1$}


In the previous section, we have introduced the operator ${\left\| \bm{R} \right\|}_n$ and proposed to measure the average $\langle {\left\| \bm{R} \right\|}_n \rangle$ to search the new transition. The global symmetry $\widetilde{\mathrm{SU(2)}}_\mathrm{global}$ is unbroken if $\langle {\left\| \bm{R} \right\|}_n \rangle \to 0$ in the infinite volume limit $V \to \infty$. On the other hands, the global symmetry $\widetilde{\mathrm{SU(2)}}_\mathrm{global}$ is broken if $\langle {\left\| \bm{R} \right\|}_n \rangle \to \mathrm{const.} > 0$ as $V \to \infty$.

When the lattice volume is finite, however, 
$\langle {\left\| \bm{R} \right\|}_n \rangle$ takes the nonzero value $\langle {\left\| \bm{R} \right\|}_n \rangle \neq 0$ even in the unbroken phase. 
In the finite volume $V$, we denote the value of $\langle {\left\| \bm{R} \right\|}_n \rangle$ in the unbroken phase by $\langle {\left\| \bm{R}_0 \right\|}_n \rangle$.



Let $\bm{r}_{0,x}$ be the random field variable on the surface $S^2$ which has the same global transformation property as $\bm{r}_x$: $\bm{r}_{0,x} \mapsto \Gamma^\dagger \bm{r}_{0,x} \Gamma$. 
Then we introduce another gauge-invariant field density $\bm{R}_0$ which is constructed in a way similar to $\bm{R}$:
\begin{align}
 \bm{R}_0 := \frac{1}{V} \sum_x \bm{r}_{0,x} = \bm{R}_0^{\dagger} \, , \quad
 \bm{R}_0 \mapsto \Gamma^\dagger \bm{R}_0 \Gamma \, .
\end{align}
We can estimate the volume dependence of $\langle {\left\| \bm{R}_0 \right\|}_n \rangle$ as $\langle {\left\| \bm{R}_0 \right\|}_n \rangle \propto \frac{1}{\sqrt{V}}$ which yields $\langle {\left\| \bm{R}_0 \right\|}_n \rangle \to 0$ as $V \to \infty
$. See Appendix B for more details.
In order to detect the spontaneous breaking of the global symmetry $\widetilde{\mathrm{SU(2)}}_\mathrm{global}$ in the finite volume $V$, therefore, we redefine the average of the \textit{gauge-invariant operator norm} $\langle {\left\| \bm{R} \right\|}_n \rangle^\mathrm{sub}$ by
\begin{align}
 \langle {\left\| \bm{R} \right\|}_n \rangle^\mathrm{sub} := \langle {\left\| \bm{R} \right\|}_n \rangle - \langle \left\| \bm{R}_0 \right\|_n \rangle \, .
\label{abr3c}
\end{align}
$\langle {\left\| \bm{R} \right\|}_n \rangle^\mathrm{sub}$ is the well-defined order parameter:
 $\langle {\left\| \bm{R} \right\|}_n \rangle^\mathrm{sub} =  0$ in the $\widetilde{\mathrm{SU(2)}}_\mathrm{global}$ unbroken phase, and $\langle {\left\| \bm{R} \right\|}_n \rangle^\mathrm{sub} \neq 0$ in the $\widetilde{\mathrm{SU(2)}}_\mathrm{global}$ broken phase.

Following the above procedure, we first measured the 1-norm of the scalar-color composite field density $\langle {\left\| \bm{R} \right\|}_1 \rangle^\mathrm{sub}$.

To determine the transition line, we observed the position at which the value of $\langle {\left\| \bm{R} \right\|}_1 \rangle^\mathrm{sub}$ as a function of the parameters $\beta$ and $\gamma$ changes from zero $\langle {\left\| \bm{R} \right\|}_1 \rangle^\mathrm{sub} = 0$ to a nonzero value $\langle {\left\| \bm{R} \right\|}_1 \rangle^\mathrm{sub} > 0$ as the results of numerical simulations.

Figure \ref{ta12} gives the measurement results of $\langle {\left\| \bm{R} \right\|}_1 \rangle^\mathrm{sub}$ in the $\beta$-$\gamma$ phase plane. The upper panels are the plots of $\langle {\left\| \bm{R} \right\|}_1 \rangle^\mathrm{sub}$ as functions of $\gamma$ on various $\beta = \mathrm{const.}$ lines, while the lower panels are the plots of $\langle {\left\| \bm{R} \right\|}_1 \rangle^\mathrm{sub}$, as functions of $\beta$ on various $\gamma = \mathrm{const.}$ lines.

Figure \ref{tad} is the transition line determined from the modified 1-norm of the scalar-color composite field density $\langle {\left\| \bm{R} \right\|}_1 \rangle^\mathrm{sub}$, by observing the results of Fig.\ref{ta12}.
It is remarkable that this new transition line divides the single Higgs-confinement region into two separated regions: the confinement region and the Higgs region.
Notice that this transition line was obtained in the gauge-independent manner, since any gauge fixing has not been imposed in the procedure of numerical simulations. 
\footnote{
Incidentally, it should be mentioned that the new transition line dividing the confinement phase in the case of the adjoint scalar field has been found quite recently in \cite{ShibataKondo23}, by performing gauge-independent numerical simulations in the similar way.
}

\begin{figure*}[!t]
\centering
\begin{subfigure}{88mm}
  \centering\includegraphics[width=85mm]{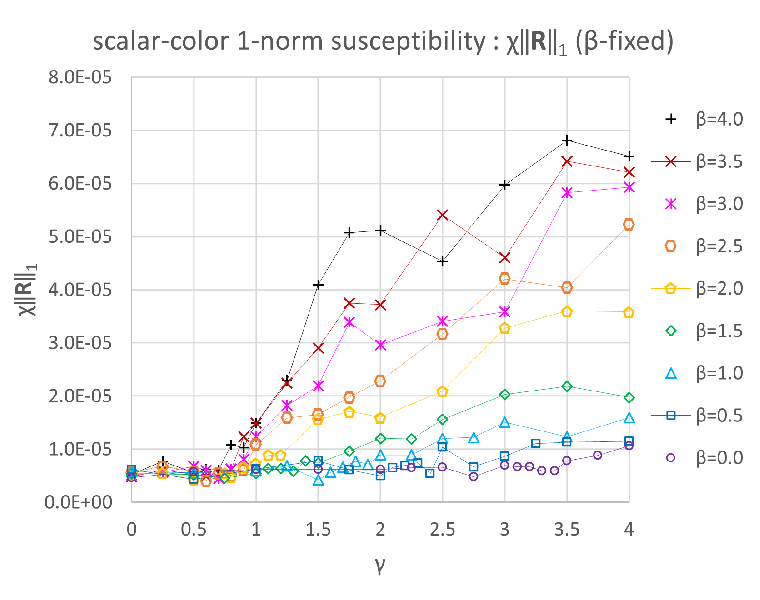}
  \label{ts1}
\end{subfigure}
\begin{subfigure}{88mm}
  \centering\includegraphics[width=85mm]{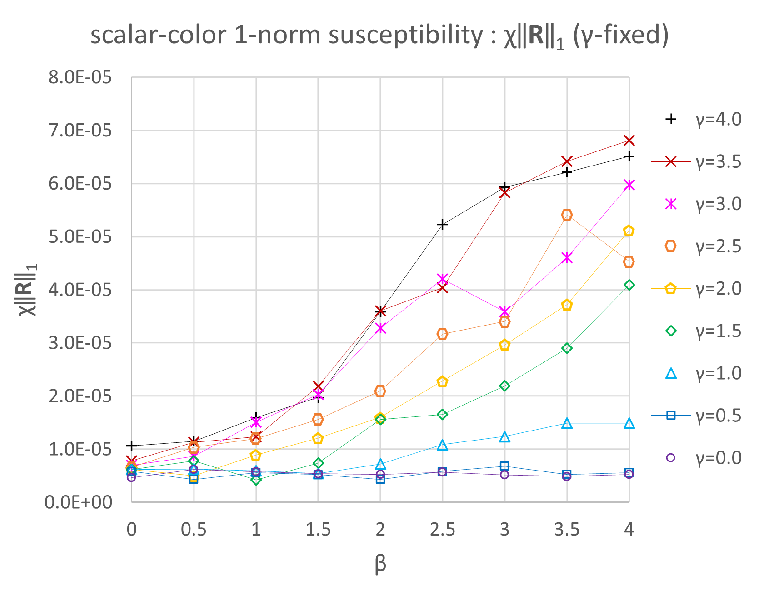}
  \label{ts2}
\end{subfigure}
\vspace{-3mm}\caption{Susceptibility ${\chi}_{ {\left\| \bm{R} \right\|}_1}$ of the 1-norm of the scalar-color composite field density on the $16^4$ lattice: (left) ${\chi}_{ {\left\| \bm{R} \right\|}_1}$ vs. $\gamma$ on various $\beta = \mathrm{const.}$ lines, (right) ${\chi}_{ {\left\| \bm{R} \right\|}_1}$ vs. $\beta$ on various $\gamma = \mathrm{const.}$ lines.}
\label{ts12}
\end{figure*}

\begin{figure*}[!t]
\centering
\begin{subfigure}{88mm}
  \centering\includegraphics[width=84mm]{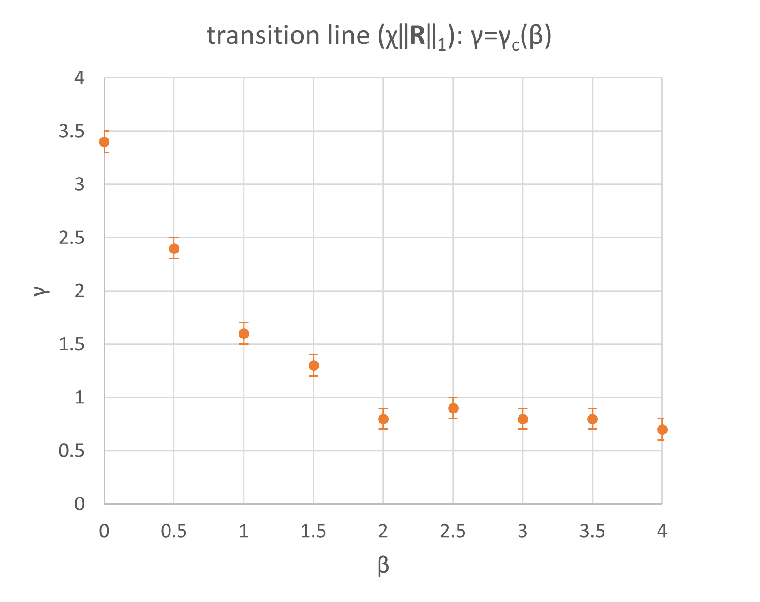}
\end{subfigure}
\vspace{-3mm}\caption{Transition lines $\gamma = {\gamma}_c (\beta)$ determined by the susceptibility ${\chi}_{ {\left\| \bm{R} \right\|}_1}$ of the 1-norm of the scalar-color composite field density ${\left\| \bm{R} \right\|}_1$ on the $16^4$ lattice.}
\label{tsd}
\end{figure*}

\subsection{Susceptibility for ${\left\| \bm{R} \right\|}_1$}

Moreover, we measured the \textit{susceptibility} (specific heat) ${\chi}_{{\left\| \bm{R} \right\|}_1} := \langle {\left\| \bm{R} \right\|}_1^2 \rangle - \langle {\left\| \bm{R} \right\|}_1 \rangle^2$ of the modified 1-norm of the scalar-color composite field density ${\left\| \bm{R} \right\|}_1$. We identified the transition lines by detecting the position at which ${\chi}_{ {\left\| \bm{R} \right\|}_1}$ changes from a constant value ${\chi}_{ {\left\| \bm{R} \right\|}_1} = {\chi}_{ {\left\| \bm{R} \right\|}_1,0} = \mathrm{const.}$ to increasing the value ${\chi}_{ {\left\| \bm{R} \right\|}_1} > {\chi}_{ {\left\| \bm{R} \right\|}_1,0} = \mathrm{const.}$.

Figure \ref{ts12} gives the measurement results of the susceptibility ${\chi}_{ {\left\| \bm{R} \right\|}_1}$ in the $\beta$-$\gamma$ phase plane. The left panel is the plots of ${\chi}_{ {\left\| \bm{R} \right\|}_1}$ as functions of $\gamma$ on various $\beta = \mathrm{const.}$ lines, while the right panel is the plots of ${\chi}_{ {\left\| \bm{R} \right\|}_1}$ as functions of $\beta$ on various $\gamma = \mathrm{const.}$ lines.

Figure \ref{tsd} is the transition line determined from the susceptibility of the modified 1-norm of the scalar-color composite field density ${\chi}_{ {\left\| \bm{R} \right\|}_1}$, by observing the results of Fig.\ref{ts12}.
This new transition line also divides the single Higgs-confinement region into the confinement region and the Higgs region.
Notice that this transition line obtained from ${\chi}_{ {\left\| \bm{R} \right\|}_1}$ agrees with that given in Fig.\ref{tad} within the errors.

\begin{figure*}[!p]
\centering
\begin{subfigure}{88mm}
  \centering\includegraphics[width=85mm]{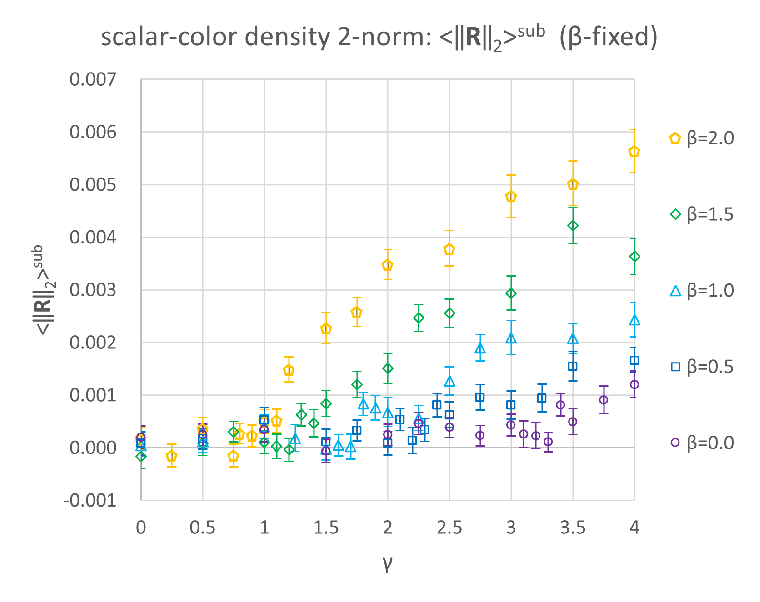}
  \label{ra1a}
\end{subfigure}
\begin{subfigure}{88mm}
  \centering\includegraphics[width=85mm]{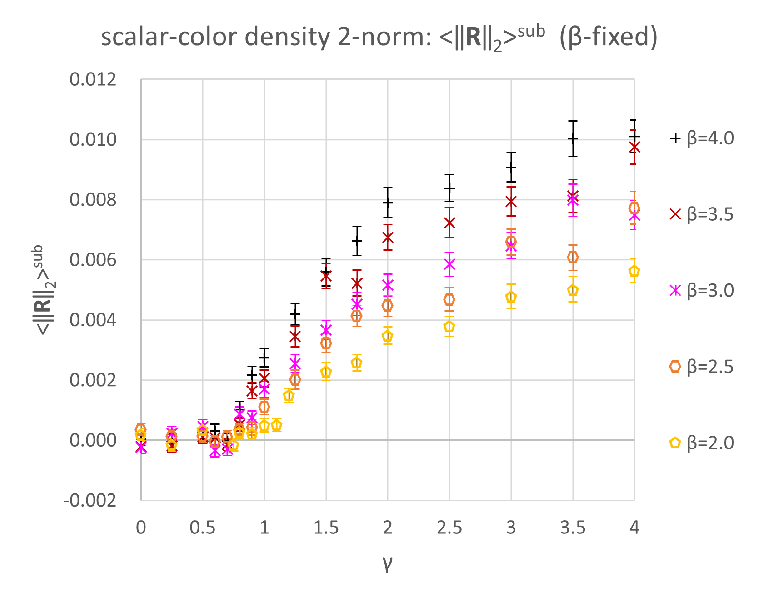}
  \label{ra1b}
\end{subfigure}\\
\begin{subfigure}{88mm}
  \centering\includegraphics[width=85mm]{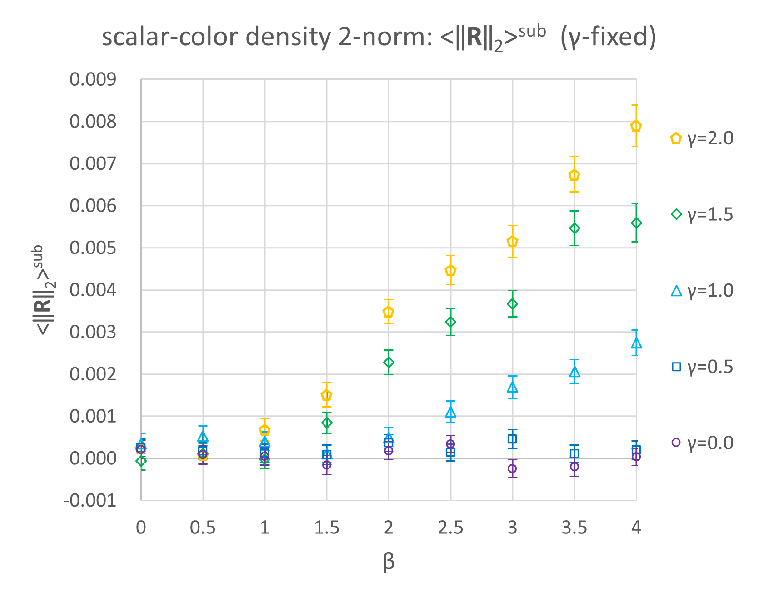}
  \label{ra2a}
\end{subfigure}
\begin{subfigure}{88mm}
  \centering\includegraphics[width=85mm]{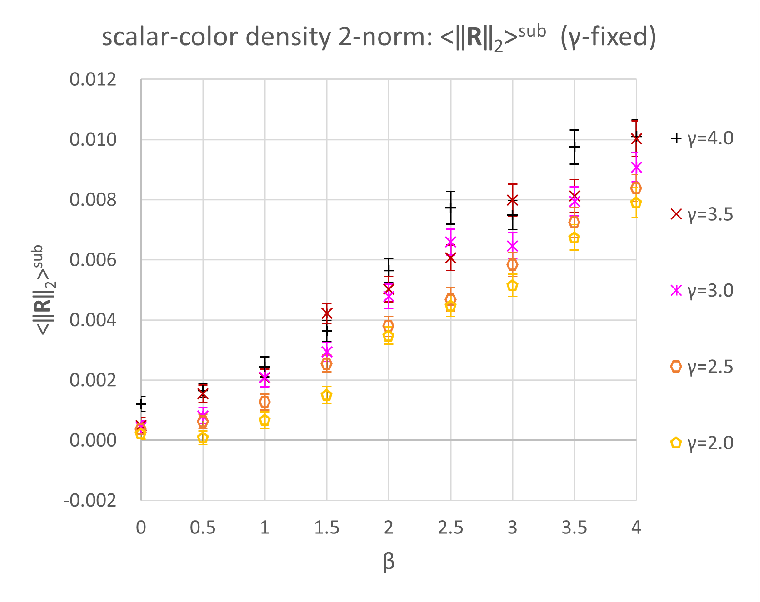}
  \label{ra2b}
\end{subfigure}
\caption{Average of the 2-norm of the scalar-color composite field density $\langle {\left\| \bm{R} \right\|}_2 \rangle^\mathrm{sub}$ on the $16^4$ lattice: (upper) $\langle {\left\| \bm{R} \right\|}_2 \rangle^\mathrm{sub}$ vs. $\gamma$ on various $\beta = \mathrm{const.}$ lines, (lower) $\langle {\left\| \bm{R} \right\|}_2 \rangle^\mathrm{sub}$ vs. $\beta$ on various $\gamma = \mathrm{const.}$ lines.}
\label{ra12}
\end{figure*}

\begin{figure*}[!p]
\centering
\begin{subfigure}{88mm}
  \centering\includegraphics[width=84mm]{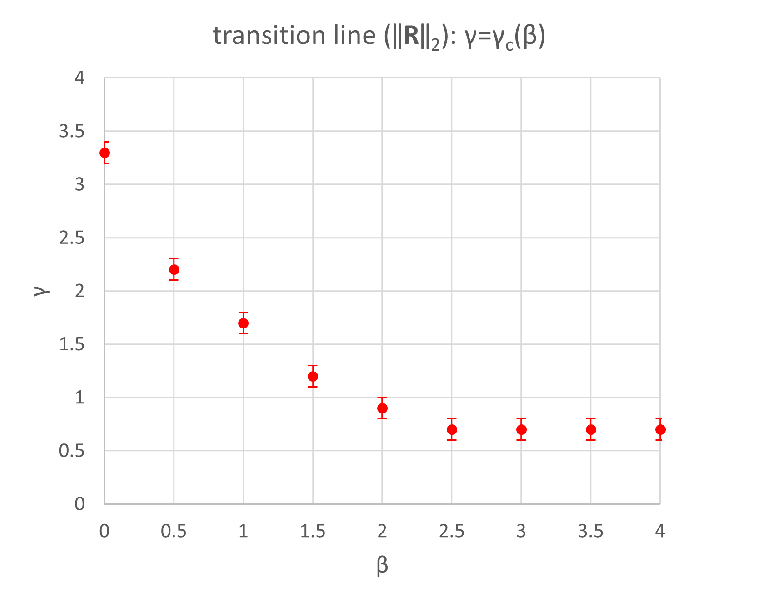}
\end{subfigure}
\caption{Transition lines $\gamma = {\gamma}_c (\beta)$ determined by the 2-norm of the scalar-color composite field density ${\left\| \bm{R} \right\|}_2$ on the $16^4$ lattice.}
\label{rad}
\end{figure*}

\subsection{Average for ${\left\| \bm{R} \right\|}_2$}

To confirm the existence of the new transition line in the phase diagram, we also measured the modified 2-norm of the scalar-color composite field density ${\left\| \bm{R} \right\|}_2$ defined in (\ref{abr3c}).  To determine the transition line, we observed the position at which $\langle {\left\| \bm{R} \right\|}_2 \rangle^\mathrm{sub}$ as a function of the parameters $\beta$ and $\gamma$ changes from zero $\langle {\left\| \bm{R} \right\|}_2 \rangle^\mathrm{sub} = 0$ to nonzero $\langle {\left\| \bm{R} \right\|}_2 \rangle^\mathrm{sub} > 0$ as the results of numerical simulations.

Figure \ref{ra12} gives the measurement results of $\langle {\left\| \bm{R} \right\|}_2 \rangle^\mathrm{sub}$ in the $\beta$-$\gamma$ phase plane. The upper panels are the plots of $\langle {\left\| \bm{R} \right\|}_2 \rangle^\mathrm{sub}$ as functions of $\gamma$ on various $\beta = \mathrm{const.}$ lines, while the lower panels are the plots of $\langle {\left\| \bm{R} \right\|}_2 \rangle^\mathrm{sub}$, as functions of $\beta$ on various $\gamma = \mathrm{const.}$ lines.

Figure \ref{rad} is the transition line determined from the modified 2-norm of the scalar-color composite field density $\langle {\left\| \bm{R} \right\|}_2 \rangle^\mathrm{sub}$ by observing the results of Fig.\ref{ra12}.
This new transition line divides the single Higgs-confinement region into two separated regions: the confinement region and the Higgs region.
Notice that this transition line was also obtained in the gauge-independent manner.
It is notable that the location of the transition line determined from the modified 2-norm of the scalar-color composite field density $\langle {\left\| \bm{R} \right\|}_2 \rangle^\mathrm{sub}$ agrees with that determined from the modified 1-norm of the scalar-color composite field density $\langle {\left\| \bm{R} \right\|}_1 \rangle^\mathrm{sub}$.

\begin{figure*}[!t]
\centering
\begin{subfigure}{88mm}
  \centering\includegraphics[width=85mm]{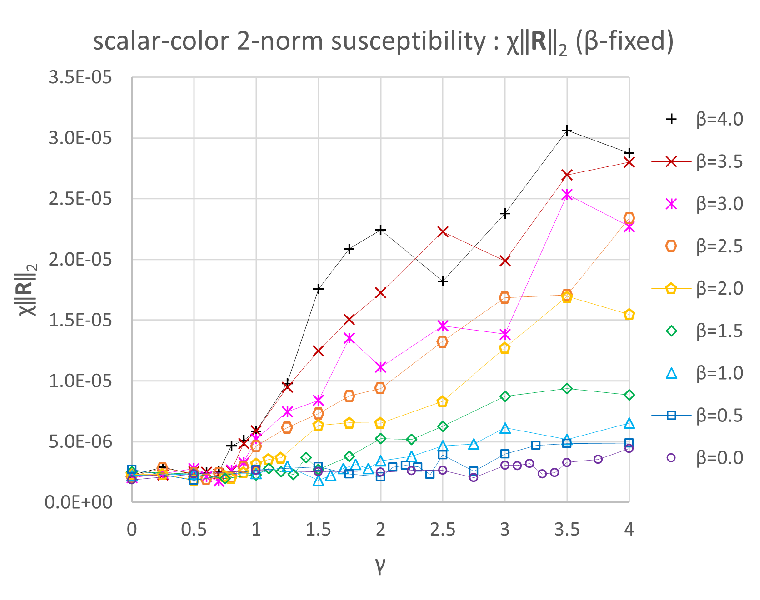}
  \label{rs1}
\end{subfigure}
\begin{subfigure}{88mm}
  \centering\includegraphics[width=85mm]{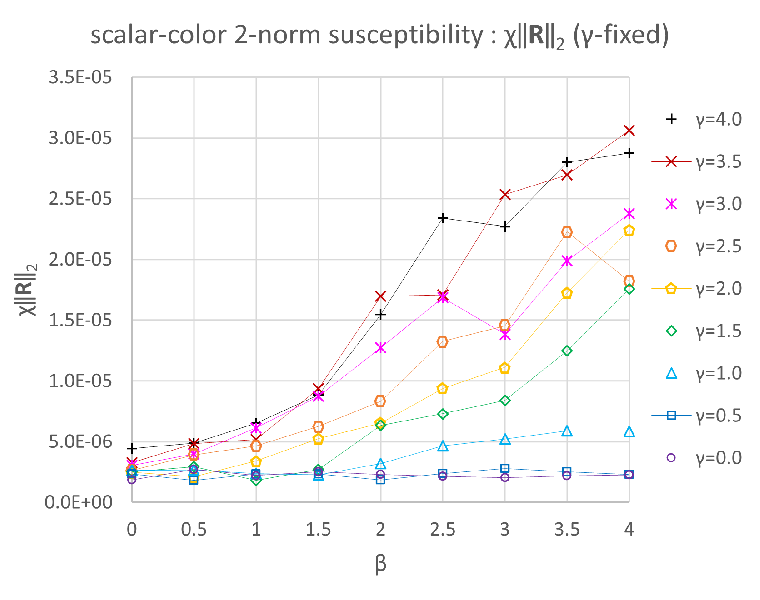}
  \label{rs2}
\end{subfigure}
\vspace{-3mm}\caption{Susceptibility ${\chi}_{ {\left\| \bm{R} \right\|}_2}$ of the 2-norm of the scalar-color composite field density ${\left\| \bm{R} \right\|}_2$ on the $16^4$ lattice: (left) ${\chi}_{ {\left\| \bm{R} \right\|}_2}$ vs. $\gamma$ on various $\beta = \mathrm{const.}$ lines, (right) ${\chi}_{ {\left\| \bm{R} \right\|}_2}$ vs. $\beta$ on various $\gamma = \mathrm{const.}$ lines.}
\label{rs12}
\end{figure*}

\begin{figure}[!t]
\centering
\begin{subfigure}{88mm}
  \centering\includegraphics[width=84mm]{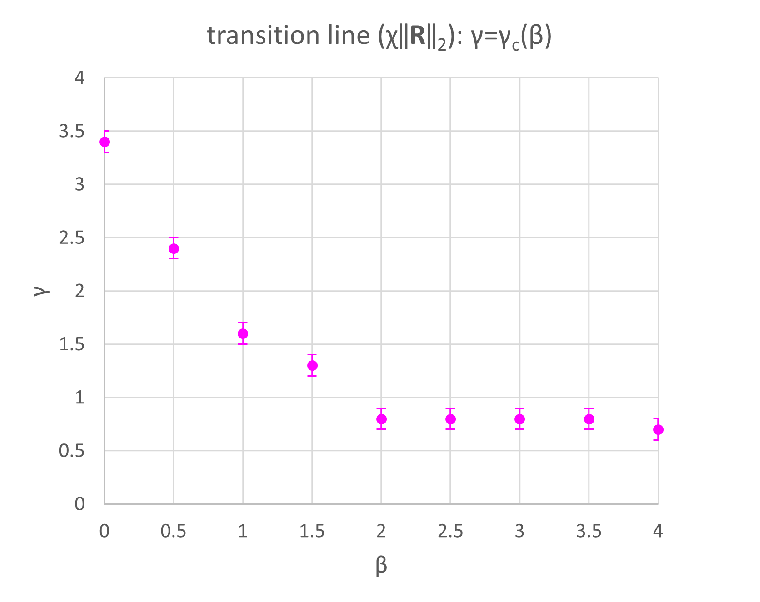}
\end{subfigure}
\vspace{-3mm}\caption{Transition lines $\gamma = {\gamma}_c (\beta)$ determined by the susceptibility ${\chi}_{ {\left\| \bm{R} \right\|}_2}$ of the 2-norm of the scalar-color composite field density ${\left\| \bm{R} \right\|}_2$ on the $16^4$ lattice.}
\label{rsd}
\end{figure}

\subsection{Susceptibility for ${\left\| \bm{R} \right\|}_2$}

Lastly, we measured the \textit{susceptibility} (specific heat) ${\chi}_{{\left\| \bm{R} \right\|}_2} := \langle {\left\| \bm{R} \right\|}_2^2 \rangle - \langle {\left\| \bm{R} \right\|}_2 \rangle^2$ of the modified 2-norm of the scalar-color composite field density ${\left\| \bm{R} \right\|}_2$. We identified the transition lines by detecting the position at which ${\chi}_{ {\left\| \bm{R} \right\|}_2}$ changes from ${\chi}_{ {\left\| \bm{R} \right\|}_2} = {\chi}_{ {\left\| \bm{R} \right\|}_2,0} = \mathrm{const.}$ to ${\chi}_{ {\left\| \bm{R} \right\|}_2} > {\chi}_{ {\left\| \bm{R} \right\|}_2,0} = \mathrm{const.}$.

Figure \ref{rs12} gives the measurement results of ${\chi}_{ {\left\| \bm{R} \right\|}_2}$ in the $\beta$-$\gamma$ phase plane. The left panel is the plots of ${\chi}_{ {\left\| \bm{R} \right\|}_2}$ as a function of $\gamma$ on various $\beta = \mathrm{const.}$ lines, while the right panel is the plots of ${\chi}_{ {\left\| \bm{R} \right\|}_2}$ as a function of $\beta$ on various $\gamma = \mathrm{const.}$ lines.

Figure \ref{rsd} is the transition line determined from the susceptibility of the modified 2-norm of the scalar-color composite field density ${\chi}_{ {\left\| \bm{R} \right\|}_2}$, by observing the results of Fig.\ref{rs12}.
This new transition line also divides the single Higgs-confinement region into the confinement region and the Higgs region.
It is remarkable that this transition line obtained from ${\chi}_{ {\left\| \bm{R} \right\|}_2}$ approximately agrees with that given in Fig.\ref{rad} within the errors.

\subsection{Volume dependence of the new transition line}

In the numerical simulations on the lattice with a finite volume, it is well known that there exists an issue coming from the finite volume effects, as reported for the gauge-scalar model in \cite{Bonati}.
For the purpose of examining the volume dependence of the new transition line, we performed the measurement of $\langle {\left\| \bm{R} \right\|}_n \rangle^\mathrm{sub}$ on $8^4$ and $16^4$ lattices.

\begin{figure}[!t]
\centering
\begin{subfigure}{88mm}
  \centering\includegraphics[width=84mm]{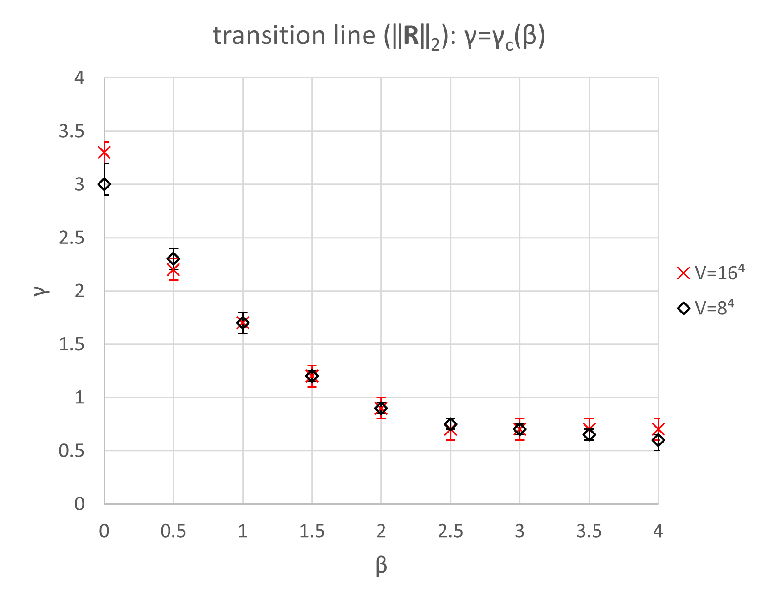}
\end{subfigure}
\vspace{-3mm}\caption{Transition lines $\gamma = {\gamma}_c (\beta)$ determined by the 2-norm of the scalar-color composite field density $\langle {\left\| \bm{R} \right\|}_2 \rangle^\mathrm{sub}$ on the $8^4$ and $16^4$ lattice.}
\label{veff}
\end{figure}

Figure \ref{veff} exhibits the transition lines determined from the modified 2-norm of the scalar-color composite field density $\langle {\left\| \bm{R} \right\|}_2 \rangle^\mathrm{sub}$ on $8^4$ and $16^4$ lattices.
These new transition lines divide the single Higgs-confinement region into two separated regions: the confinement region and the Higgs region.
The location of the transition line determined on the $16^4$ lattice is shifted upward in $\gamma$, compared with the transition line determined on the $8^4$ lattice.
In fact, the small $\beta$ region, especially, $\beta = 0$ case is very sensitive to the finite volume effect. Therefore, we cannot conclude whether the transition line reaches $\beta = 0$ line at a certain finite value of $\gamma$ or not.
\footnote{
In our simulations, indeed, we adopted the very small positive value $\beta = 10^{-14}$ and $\gamma = 10^{-14}$ instead of $\beta = 0$ and $\gamma = 0$ which means  excluding the gauge action and the scalar action respectively from the beginning.
}

However, the results do not excludes the transition line which terminates at the horizontal axis $\gamma=\infty$, although our data of numerical simulations available are not sufficient to conclude the precise position of the transition line. 
Indeed, it is shown in Appendix A that the spontaneous symmetry breaking of the global symmetry $\widetilde{\mathrm{SU(2)}}_\mathrm{global}$ can occur at $\gamma=\infty$ based on another reformulation of this model.
See Figure \ref{2ph} for the schematic phase diagram.

\FloatBarrier
\section{Understanding the new phase structure}

\begin{figure}[!t]
\centering
\begin{subfigure}{75mm}
  \centering\includegraphics[width=75mm]{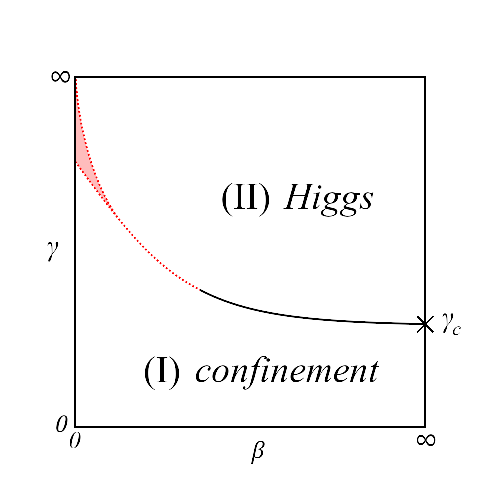}
\end{subfigure}
\caption{The schematic phase diagram: (I) confinement phase and (II) Higgs phase. 
The red area describes the possible locations of the new transition line due to finite volume effects.
}
\label{2ph}
\end{figure}

According to our numerical simulations for the deformed theory, the phase diagram is divided into confinement phase (I) $\gamma<\gamma_c(\beta)$ ($\langle {\left\| \bm{R} \right\|}_2 \rangle^\mathrm{sub} = 0$) and Higgs phase (II) $\gamma>\gamma_c(\beta)$ ($\langle {\left\| \bm{R} \right\|}_2 \rangle^\mathrm{sub} \neq 0$) as shown schematically in Fig.\ref{2ph}.

In what follows we discuss why the above phase structure does not contradict with the conventional wisdoms and how the respective phase is characterized from the physical point of view. 

First, we discuss why the above phase structure does not contradict with the conventional wisdoms.

According to the Elitzur theorem\cite{Elitzur75}, the local gauge symmetry cannot be spontaneously broken, or there are no (local) order parameters for detecting the spontaneous breaking of local gauge symmetry. On the other hand, if there exists a global symmetry, there also exists a corresponding local order parameter.
However, there can exist no local order parameter for the symmetry which enable to discriminate between the confinement phase and the Higgs phase \cite{Maas19}.
Therefore, the new transition detected by the nonlocal order parameter $\langle {\left\| \bm{R} \right\|}_n \rangle^\mathrm{sub}$ defined by (\ref{abr3c}) is not related to the original global symmetry $\mathrm{SU(2)}_\mathrm{global}$ and it is not identified with the actual thermodynamic transition line.
In order to show the separation between confinement and Higgs regions, we must temporarily deform the theory such that it has a certain nonlocal character which invalidates the OSFS theorem.

(i) Confinement phase (I) and Higgs phase (II) can be respectively characterized by the absence or presence of spontaneous breaking of the global symmetry $\widetilde{\mathrm{SU(2)}}_\mathrm{global}$.

Notice that $\bm{R}$ is a Hermitian matrix. Therefore, $\bm{R}$ can be diagonalized by a unitary matrix and can be expressed using the real-valued eigenvalues $\lambda_\pm$ defined in (\ref{eigenv}) as 
\begin{align}
 &\bm{R}
  = \begin{pmatrix} R^3 & R^1+iR^2 \\ R^1-iR^2 & -R^3 \end{pmatrix}
  = \Gamma_* \begin{pmatrix} \lambda_+ & 0 \\ 0 & \lambda_- \end{pmatrix} \Gamma_{*}^{ \dagger} \, , \notag\\
 &\Gamma_* \in \widetilde{\mathrm{SU(2)}}_\mathrm{global} \, ,
\end{align}
where $\Gamma_*$ represents a certain matrix of $\widetilde{\mathrm{SU(2)}}_\mathrm{global}$ which realizes the diagonalization.
To obtain the nonvanishing average avoiding the cancellations between $\lambda_+$ and $\lambda_-$ ($\lambda_+=-\lambda_->0$), we use only $\lambda_+>0$. 

Higgs phase (II) is characterized by 
$\langle {\left\| \bm{R} \right\|}_2 \rangle^\mathrm{sub} \neq 0$.
In this phase, a specific rotation matrix $\Gamma_* \in \widetilde{\mathrm{SU(2)}}_\mathrm{global}$ is chosen to realize the diagonalization of the matrix $\bm{R}$ with nonzero eigenvalue $\lambda = \lambda_\pm \neq 0$.
Therefore, Higgs phase (II) is interpreted as an ordered phase with the spontaneously broken global symmetry $\widetilde{\mathrm{SU(2)}}_\mathrm{global}$.

Confinement phase (I) is characterized by 
$\langle {\left\| \bm{R} \right\|}_2 \rangle^\mathrm{sub} = 0$.
In this phase, any specific rotation matrix $\Gamma_*$ is not needed. Therefore, confinement phase (I) is interpreted as a disordered phase with the unbroken global symmetry $\widetilde{\mathrm{SU(2)}}_\mathrm{global}$.

Notice that the above argument has nothing to do with the local gauge symmetry $\mathrm{SU(2)}_\mathrm{local}$ for $\bm{R}$. Therefore, the local symmetry $\mathrm{SU(2)}_\mathrm{local}$ is unbroken in both phases. 
Therefore, confinement phase (I) is the phase where both the local gauge symmetry $\mathrm{SU(2)}_\mathrm{local}$ and the global symmetry $\widetilde{\mathrm{SU(2)}}_\mathrm{global}$ being  unbroken ($\langle {\left\| \bm{R} \right\|}_2 \rangle^\mathrm{sub} = 0$), while Higgs phase (II) is the phase where the local gauge symmetry $\mathrm{SU(2)}_\mathrm{local}$ is unbroken but the global symmetry $\widetilde{\mathrm{SU(2)}}_\mathrm{global}$ is spontaneously broken ($\langle {\left\| \bm{R} \right\|}_2 \rangle^\mathrm{sub} \neq 0$).

(ii) 
The existence of a new transition line we found does not contradict with the OSFS analyticity theorem. 

The gauge-scalar model discussed in Osterwalder-Seiler\cite{OsterwalderSeiler78} and Fradkin-Shenker\cite{FradkinShenker79} has the same symmetry as that of our original model, although the symmetry is realized nonlinearly in Osterwalder-Seiler model, while it is realized linearly in our original model. Therefore, the OSFS theorem is applicable to the original gauge-scalar model.

However, the operator such as the intrinsically nonlocal operator $\bm{R}$ in the deformed theory (\ref{exvev2}) is not supposed in the proof of the OSFS theorem which states the analyticity between confinement and Higgs regions in the phase plane $(\beta,\gamma)$ of the original theory.
 $\bm{R}$ includes the color-direction field obtained according to (\ref{redc2}) through the reduction procedure which involves the gauge field configurations over the whole lattice.
In the proof of analyticity \cite{OsterwalderSeiler78}, a convergent \textit{cluster expansion} used for the expectation value of a local operator is valid only if the operator has a finite support.
Therefore, the OSFS theorem is not applicable to $\bm{R}$ in the deformed theory (\ref{exvev2}).
Thus, the existence of the new transition line detected by $\bm{R}$ does not contradict with the OSFS theorem.

(iii) 
The massless Nambu-Goldstone particles do not appear in the Higgs phase even if the continuous global symmetry $\widetilde{\mathrm{SU(2)}}_\mathrm{global}$ is spontaneously broken in the deformed theory (\ref{exvev2}).
According to the conventional understanding, if the scalar field $\hat{\Theta}_x$ acquires a nonvanishing vacuum expectation value (VEV) $\langle \hat{\Theta}_x \rangle=\frac{v}{\sqrt{2}} \bm{1}$ in the unitary gauge, the symmetry $\mathrm{SU(2)}_\mathrm{local} \times \widetilde{\mathrm{SU(2)}}_\mathrm{global}$ of the action is spontaneously broken down to a diagonal global subgroup $\mathrm{SU(2)}_\mathrm{diag}$: $\mathrm{SU(2)}_\mathrm{local} \times \widetilde{\mathrm{SU(2)}}_\mathrm{global} \to \mathrm{SU(2)}_\mathrm{diag}$ 
such that the VEV of $\hat{\Theta}_x$ is preserved under the transformation 
${\Omega}_x = \Gamma^{\dagger} = G$:
\begin{align}
  U_{x,\mu} &\mapsto {\Omega}_x U_{x,\mu} {\Omega}_{x+\mu}^{\dagger}   \Longrightarrow U_{x,\mu}  \mapsto G U_{x,\mu} G^{\dagger} , \notag\\
  \hat{\Theta}_x &\mapsto {\Omega}_x \hat{\Theta}_x \Gamma \hspace{8.25mm} \Longrightarrow \hspace{2.25mm} \hat{\Theta}_x \mapsto G \hat{\Theta}_x G^{\dagger}   .
\end{align}
%
%
In order to introduce the new gauge-invariant composite operator $\bm{r}_x := \hat{\Theta}_x^{\dagger} \bm{n}_x \hat{\Theta}_x$ to detect the spontaneous symmetry breaking, we need to obtain the color-direction field $\bm{n}_x$. However, the color-direction field $\bm{n}_x$ is obtained by minimizing the reduction functional which involves the  gauge configurations $\{ U_{x,\mu} \}$ over the whole lattice and given as an integral over the whole the lattice volume.
Therefore, the resulting color-direction field $\bm{n}_x$ is intrinsically nonlocal despite its appearance, which violates one of the assumptions, i.e.,  locality in proving the Nambu-Goldstone theorem.

This is the reason why there are no massless particles (gapless excitations) in the Higgs phase, although the continuous global symmetry $\widetilde{\mathrm{SU(2)}}_\mathrm{global}$ is spontaneously broken.


Next, we discuss how the respective phase is characterized from the physical point of view. 

(i) 
First, we consider the confinement phase (I) $\gamma<\gamma_c(\beta)$ below the new critical line $\gamma=\gamma_c(\beta)$ where $\langle {\left\| \bm{R} \right\|}_2 \rangle^\mathrm{sub}$ takes the vanishing value $\langle {\left\| \bm{R} \right\|}_2 \rangle^\mathrm{sub} = 0$.

In the limit $\gamma \to 0$, especially, the SU(2) gauge-scalar model reduces to the pure compact SU(2) gauge model which is expected to have a single confinement phase with no phase transition and has a mass gap on the whole $\beta$ axis in four spacetime dimensions \cite{Creutz82}. 

Confinement is expected to occur due to vacuum condensations of appropriate topological defects, e.g., magnetic monopoles for non-Abelian gauge theory \cite{dualsuper}.  Here such topological defect should be carefully defined gauge-independently using the gauge-invariant method, which is actually realized by extending the gauge-covariant decomposition of the gauge field, see \cite{KKSS15} for a review.  

Even in a finite $\gamma$ region: $0 < \gamma<\gamma_c(\beta)$, the effect of the scalar field would be relatively small and confinement would occur in the way similar to the pure SU(2) gauge model, which we call confinement phase (I) from the belief that the original gauge symmetry SU(2) would be kept unbroken and not spontaneously broken. 

Confinement phase (I)  is regarded as a disordered phase in which all the symmetries are restored. 
In confinement phase (I), therefore, the color-direction field $\bm{n}_{x}$ takes various possible directions with no specific direction (isotropic configuration) in color space.
This can be estimated through $\langle {\left\| \bm{R} \right\|}_2 \rangle^\mathrm{sub}$ in relation to the direction of the fundamental scalar field $\hat{\Theta}_{x}$. The vanishing of the average $\langle {\left\| \bm{R} \right\|}_2 \rangle^\mathrm{sub} = 0$ is caused by the very small correlation between the color-direction field $\bm{n}_x$ and the fundamental scalar field $\hat{\Theta}_x$.
Therefore, a single phase with a mass gap is expected to exist in the region (I). 
The gauge fields become massive due to self-interactions among themselves.

(ii)
Next, we consider the Higgs phase (II) $\gamma>\gamma_c(\beta)$ above the new critical line where $\langle {\left\| \bm{R} \right\|}_2 \rangle^\mathrm{sub}$ takes the nonvanishing value $\langle {\left\| \bm{R} \right\|}_2 \rangle^\mathrm{sub} \neq 0$.

In the Higgs phase (II)  the gauge fields become massive due to different mechanism from that in the confinement phase (I). 
According to the conventional BEH mechanism, this phenomenon is understood as a consequence of the (complete) spontaneous symmetry breaking $\mathrm{SU(2)} \to \{ \bm{1} \}$. 
Note that the \textit{gauge-independent description of the BEH mechanism} \cite{Kondo18} provides the new interpretation without introducing the spontaneous gauge symmetry breaking. 
Therefore, the Higgs phase (II) with massive gauge fields is also expected to exist  due to the absence of massless gauge mode throughout this phase.
In the limit $\gamma \to \infty$, all components of the gauge field become infinitely heavy and decouple from the physical modes and there is no remaining massless diagonal gauge field unlike the lattice gauge-scalar model with the adjoint scalar field \cite{ShibataKondo23}.  

This phase is characterized by the nonvanishing value $\langle {\left\| \bm{R} \right\|}_2 \rangle^\mathrm{sub} \neq 0$, which means that the color-direction field $\bm{n}_{x}$ correlates strongly with the given fundamental scalar field $\hat{\Theta}_x$ which tends to align to an arbitrary but a specific direction as expected from the spontaneous symmetry breaking in an ordered phase.
The Higgs phase (II) is regarded as an ordered phase in which the color-direction field $\bm{n}_x$ takes the anisotropic configuration in color space together with the  fundamental scalar field $\hat{\Theta}_x$.

\section{Conclusions and discussions}

In this paper, we reexamined the phase structure of the lattice SU(2) gauge-scalar model with the scalar field in the fundamental representation of the gauge group by introducing the new type of gauge-invariant operators. 
According to the preceding studies \cite{FradkinShenker79,OsterwalderSeiler78,LangRebbiVirasoro81,Bonati}, this model has a single confinement-Higgs phase composed of analytically continued confinement and Higgs subregions, and therefore there are no thermodynamic phase transitions between the two regions.

We constructed gauge-invariant composite operators composed of the fundamental scalar field and the {color-direction field} constructed from the gauge field which can be obtained from change of field variables \cite{KKSS15} based on the gauge-covariant decomposition of the gauge field due to Cho-Duan-Ge-Shabanov \cite{Cho8081,DuanGe79,Shabanov99} and Faddeev-Niemi \cite{FaddeevNiemi9907}. 
We found the gauge-independent separation between the confinement phase and the Higgs phase without any specific gauge fixing.

We performed the gauge-fixing-free numerical simulations. We reproduced the conventional thermodynamic transition line in the weak gauge coupling region \cite{LangRebbiVirasoro81,Bonati}. Moreover, we confirmed that there exists a new transition line divides a single confinement-Higgs phase into the confinement phase and the Higgs phase in the strong gauge coupling region.
We provided a possible physical interpretation of the new transition and the resulting separated phases as a symmetric and spontaneously broken realization of a global continuous symmetry $\widetilde{\mathrm{SU(2)}}_\mathrm{global}$ of the deformed theory which should be discriminated from the global symmetry $\mathrm{SU(2)}_\mathrm{global}$ of the original theory. 
Notice that the $\mathrm{SU(2)}_\mathrm{global}$ symmetry of the original theory is not broken anywhere in the phase diagram, in accord with the OSFS theorem. In other words, the new transition line separating confinement and Higgs phases appears only when the theory is deformed so as to have a certain nonlocality which invalidates the OSFS theorem.

Finally, we can say something about confinement or deconfinement in this model. 
In the confinement phase (I)  there would occur magnetic monopole condensations which will play the dominant role in realizing quark confinement based on the dual superconductor picture. 
In fact, the magnetic monopole can be constructed only from the gauge degrees of freedom through the color-direction field and the magnetic monopole dominance in quark confinement has been confirmed in the pure gauge case in the gauge-invariant way \cite{KKSS15}. 
In the Higgs phase (II), on the other hand, there would be no magnetic-monopole condensations and confinement would not occur. This issue is to be clarified in the next work.
In subsequent papers, we will give more detailed theoretical and numerical investigations to confirm the new transition line, then discuss physical implications of the new transition including confinement or deconfinement in view of \cite{FMS81,tHooft80} and also the extension to the gauge-fundamental scalar model with the gauge group SU(3) with a different global symmetry \cite{Maas17}.



\appendix
\section{ANOTHER FORMULATION FOR THE GAUGE-INDEPENDENT BEH MECHANISM AND SPONTANEOUS GLOBAL SYMMETRY BREAKING}

In this Appendix, we give another formulation of the SU(2) gauge-scalar model with a single radially-fixed fundamental scalar field. 
It turns out that the new formulation enables to give the gauge-independent description of the BEH mechanism and its relation to the spontaneous global symmetry breaking. 

First of all, we introduce the new link variable $\tilde{W}_{x,\mu}$ by 
\footnote{
The authors would like to thank Professor Jun Nishimura for calling our attention to the importance of the new link variable to understand some aspects of our results, and subsequent discussions on the related issues. 
}
\begin{align}
 &\tilde{W}_{x,\mu} :=  {\hat{\Theta}}_x^{\dagger} U_{x,\mu} {\hat{\Theta}}_{x+\mu} \notag\\
 &\Leftrightarrow \ U_{x,\mu} = {\hat{\Theta}}_x \tilde{W}_{x,\mu} {\hat{\Theta}}_{x+\mu}^{\dagger} \,.
\end{align}
This link variable $\tilde{W}_{x,\mu}$ is gauge invariant:
\begin{align}
  \tilde{W}_{x,\mu} &\mapsto \tilde{W}_{x,\mu}^{\prime} = \tilde{W}_{x,\mu} 
\end{align} 
under the local gauge transformation ${\Omega}_x \in \mathrm{SU(2)}_\mathrm{local}$ for the link variable $U_{x,\mu}$
and the site variable ${\hat{\Theta}}_x$ as the fundamental scalar field: 
\begin{align}
 U_{x,\mu}
  &\mapsto U_{x,\mu}^{\prime} = {\Omega}_x U_{x,\mu} {\Omega}_{x+\mu}^{\dagger} \, , \notag\\
 \hat{\Theta}_x
  &\mapsto \hat{\Theta}_x^{\prime} = {\Omega}_x \hat{\Theta}_x \, , \quad {\Omega}_x \in \mathrm{SU(2)}_\mathrm{local} \,. 
\end{align} 
Then we find that the gauge-invariant lattice action of the gauge-scalar model with the fundamental scalar field can be rewritten in terms of the new gauge-invariant variable $\tilde{W}_{x,\mu}$ alone:
\begin{align}
 &S[U,\hat{\Theta}]
  = S_G [U] + S_H [U,\hat{\Theta}]
  = \tilde{S}_G [\tilde{W}] + \tilde{S}_H [\tilde{W}] \,, \\
 &S_G [U]
  = \tilde{S}_G [\tilde{W}] \notag\\
 &= \frac{\beta}{2} \sum_{x,\mu>\nu} \re \tr \left( \mathbf{1} - \tilde{W}_{x,\mu} \tilde{W}_{x+\mu,\nu} \tilde{W}_{x+\nu,\mu}^{\dagger} \tilde{W}_{x,\nu}^{\dagger} \right) , \\
 &S_H [U,\hat{\Theta}]
  = \tilde{S}_H [\tilde{W}] \notag\\
 &= \frac{\gamma}{2} \sum_{x,\mu} \re \tr \left( \mathbf{1} - \tilde{W}_{x,\mu} \right) .
\end{align}
Moreover, we pay attention to the integration measure. 
It is shown that  
\begin{align}
 \prod_{x,\mu} dU_{x,\mu} \prod_{x} d\hat{\Theta}_{x} 
 = \prod_{x,\mu} d\tilde{W}_{x,\mu} \prod_{x} d\hat{\Theta}_{x} \,,
\end{align} 
which follows from the fact that the Jacobian associated with change of variables $(U_{x,\mu},\hat{\Theta}_{x}) \to (\tilde{W}_{x,\mu},\hat{\Theta}_{x})$ is essentially equal to one. 

Thus, the two theories: the original theory of the action $S_G [U] + S_H [U,\hat{\Theta}]$ with the measure $\prod_{x,\mu} dU_{x,\mu} \prod_{x} d\hat{\Theta}_{x}$ and the new theory of the action $\tilde{S}_G [\tilde{W}] + \tilde{S}_H [\tilde{W}]$ with the measure $\prod_{x,\mu} d\tilde{W}_{x,\mu} \prod_{x} d\hat{\Theta}_{x}$ are equivalent.  

The characteristic properties and advantages of the new formulation are as follows.
\footnote{
Numerical simulations of SU(2) gauge-fundamental scalar model based on the new formulation have been done by Montvay \cite{Montvay}.
}

\noindent
(1) [local gauge symmetry $\mathrm{SU(2)}_\mathrm{local}$ and gauge-invariant massive  gauge boson]

The link variable $\tilde{W}_{x,\mu}$ is the lattice version of the \textit{gauge-invariant massive  gauge boson field} $\tilde{\mathscr{W}}_\mu$ introduced  in the continuum formulation as Eq. (81) in \cite{Kondo18}:
\begin{align}
 &\tilde{\mathscr{W}}_\mu(x)
  := ig^{-1} \hat{\Theta}(x)^\dagger {D}_{\mu}[\mathscr{A}] \hat{\Theta}(x) \notag\\
  &= ig^{-1} \hat{\Theta}(x)^\dagger (\partial_{\mu} \hat{\Theta}(x) -ig \mathscr{A}_{\mu}(x)\hat{\Theta}(x)) \,.
\end{align}
Indeed, $\tilde{S}_H [\tilde{W}]$ is reduced to the gauge-invariant mass term $\frac12 M_W^2 \tilde{\mathscr{W}}_\mu(x)\tilde{\mathscr{W}}^\mu(x)$ for the gauge boson field $\tilde{\mathscr{W}}_\mu(x)$ in the continuum limit as shown by expanding the new link variable $\tilde{W}_{x,\mu}=\exp (ig \epsilon \tilde{\mathscr{W}}_\mu(x))$  with a lattice spacing $\epsilon$ in powers of the Lie-algebra valued field $\tilde{\mathscr{W}}_\mu(x)$. Hence, $\gamma$ is set to be proportional to the bare gauge boson mass squared $M_W^2$: $\gamma \propto M_W^2$. 

Therefore, the action of the SU(2) gauge-scalar model can be completely written in terms of the gauge-invariant massive modes if the  scalar field is fundamental. 
According to the conventional BEH mechanism, the fundamental scalar field causes complete spontaneous breaking of the original gauge symmetry $\mathrm{SU(2)}_\mathrm{local}$ and thereby all the components of the gauge field become massive by absorbing all (would-be) massless Nambu-Goldstone particles appearing according to the Nambu-Goldstone theorem associated to the spontaneous symmetry breaking.  
This fact suggests that the new formulation is more suitable than the original one to discuss the Higgs phase in contrast to the confinement phase which can be well described in the original formulation.
The new formulation can give the gauge-invariant (independent) description of the BEH mechanism in the case of the fundamental scalar field initiated in \cite{Kondo18}. 

This should be compared with the SU(2) gauge-scalar model with the adjoint scalar field which exhibits the partial spontaneous symmetry breaking in the sense that the original gauge symmetry SU(2) is broken into the nontrivial subgroup U(1) which corresponds to the massless gauge mode. Therefore, in this model the theory cannot be rewritten in terms of the massive modes alone even after the BEH phenomenon occurs \cite{Kondo16}.  

\noindent
(2) [global symmetry $\Gamma \in \mathrm{SU(2)}_\mathrm{global}$]

Although the link variable $\tilde{W}_{x,\mu}$ is gauge-invariant, 
it transforms under the global transformation $\Gamma \in \mathrm{SU(2)}_\mathrm{global}$ for the site variable ${\hat{\Theta}}_x$ according to the adjoint representation:  
\begin{align}
 &\hat{\Theta}_x \mapsto \hat{\Theta}_x^{\prime} = \hat{\Theta}_x \Gamma
  \ \Rightarrow \ \tilde{W}_{x,\mu} \mapsto \tilde{W}_{x,\mu}^{\prime} = \Gamma^\dagger \tilde{W}_{x,\mu} \Gamma \,, \notag\\
 &\Gamma \in \mathrm{SU(2)}_\mathrm{global} \,.
\end{align}
Under this global transformation, the original lattice action of the gauge-scalar model with the fundamental scalar field is invariant. Therefore, this is also the case for the new action: 
\begin{align}
 &\tilde{S}_G [\tilde{W}] 
 = \frac{\beta}{2} \sum_{x,\mu>\nu} \re \tr \left( \mathbf{1} - \tilde{W}_{x,\mu} \tilde{W}_{x+\mu,\nu} \tilde{W}_{x+\nu,\mu}^{\dagger} \tilde{W}_{x,\nu}^{\dagger} \right) \notag\\
  &\Rightarrow S_G [\tilde{W}^{\prime}]=S_G [\tilde{W}] \, , \\
 &\tilde{S}_H [\tilde{W}] 
   = \frac{\gamma}{2} \sum_{x,\mu} \re \tr \left( \mathbf{1} - \tilde{W}_{x,\mu} \right) \notag\\
  &\Rightarrow \tilde{S}_H [\tilde{W}^{\prime}]=\tilde{S}_H [\tilde{W}] \, .
\end{align}
In the conventional standpoint, this can be understood that the original gauge symmetry is explicitly broken by the mass term and the symmetry of the theory reduces to the global $\mathrm{SU(2)}_\mathrm{global}$ symmetry with no local gauge symmetry, since all the gauge fields become massive and there are no massless mode which respect the local gauge symmetry.

\noindent
(3) [no thermodynamic phase transition in the phase diagram]

At $\gamma=\infty$, all the variables $\tilde{W}_{x,\mu}$ are fixed: $\tilde{W}_{x,\mu}=\bm{1}$. 
Therefore, the theory loses the $\beta$ dependence and hence no transition occurs  at $\gamma=\infty$. 
This result is reasonable because the gauge bosons have the infinite mass at $\gamma=\infty$ and decouple from the spectrum and the theory becomes trivial. 

At $\beta=0$, namely, on the $\gamma$ axis, the variables $\tilde{W}_{x,\mu}$ on the links become mutually independent. Therefore, the average of the operator obtained by product of the operators defined on the respective link show no ``thermodynamic'' phase transition as far as we use the operator with the support consisting of a finite number of links or sites on the lattice. 
This results is consistent with the Osterwalder-Seiler-Fradkin-Shenker (OSFS) result.  Indeed, they adopted the unitary gauge ${\hat{\Theta}}_x=\bm{1}$ to show the analyticity.  
The theory at $\beta=0$ represents an ultralocal theory of the gauge boson which has no kinetic term with only a mass term. 

\noindent
(4) [new reduction and new color-direction field to study the spontaneous global symmetry]

In the new theory, the reduction functional is rewritten into 
\begin{align}
 F_\mathrm{red} [\bm{n} ; U]
   =& \sum_{x,\mu} \tr \left( \bm{1} - \bm{n}_x U_{x,\mu} \bm{n}_{x+\mu} U_{x,\mu}^{\dagger} \right) \nonumber\\
   =& \sum_{x,\mu} \tr \left( \bm{1} - \tilde{\bm{n}}_x \tilde{W}_{x,\mu} \tilde{\bm{n}}_{x+\mu} \tilde{W}_{x,\mu}^{\dagger} \right) \notag\\
   :=& \tilde{F}_\mathrm{red} [\tilde{\bm{n}} ; \tilde{W}] ,
\label{red2}
\end{align}
where we introduced a new color-direction field $\tilde{\bm{n}}_x$ defined by
\begin{align}
 &\tilde{\bm{n}}_x := {\hat{\Theta}}_x^{\dagger} \bm{n}_{x} {\hat{\Theta}}_{x} \equiv \bm{r}_x .
\end{align}
It should be remarked that the new color-direction field $\tilde{\bm{n}}_x$ is invariant under the local gauge transformation, while it transforms according to the adjoint representation under the global transformation (\ref{utn}), 
\begin{align}
 \tilde{\bm{n}}_x \mapsto \tilde{\bm{n}}_x^{\prime} = \Gamma^\dagger \tilde{\bm{n}}_x \Gamma \, , \quad \Gamma \in \widetilde{\mathrm{SU(2)}}_\mathrm{global} \, ,
\end{align}
because it is identical to the local gauge-invariant  \textit{scalar-color composite field} $\bm{r}_x$ in the original formulation:
$
 \bm{r}_x := \hat{\Theta}_x^{\dagger} \bm{n}_x \hat{\Theta}_x = \bm{r}_x^\dagger 
$.
Therefore, the new color-direction field configuration $\{ \tilde{\bm{n}}_x \}$ is determined from a given massive gauge boson field configuration $\{ \tilde{W}_{x,\mu} \}$ obtained by the new formulation by minimizing the reduction functional $\tilde{F}_\mathrm{red} [\tilde{\bm{n}} ; \tilde{W}]$ under the (enlarged) gauge transformations. Notice that the new reduction functional $\tilde{F}_\mathrm{red} [\tilde{\bm{n}} ; \tilde{W}]$ respects the global symmetry $\widetilde{\mathrm{SU(2)}}_\mathrm{global}$. 

We examine the spontaneous breaking of the global symmetry $\widetilde{\mathrm{SU(2)}}_\mathrm{global}$ which is signaled by the nonlocal operator $\bar{\tilde{\bm{n}}}$ constructed from the new color-direction field $\tilde{\bm{n}}_x$ associated with the massive gauge field $\tilde{W}_{x,\mu}$:
\begin{align}
 &\bar{\tilde{\bm{n}}} := \frac{1}{V} \sum_x \tilde{\bm{n}}_x  = \frac{1}{V} \sum_x \hat{\Theta}_x^{\dagger} \bm{n}_x \hat{\Theta}_x = \bar{\tilde{\bm{n}}}^\dagger \, , \notag\\
 &\bar{\tilde{\bm{n}}} \mapsto \bar{\tilde{\bm{n}}}^{\prime} = \Gamma^\dagger \bar{\tilde{\bm{n}}} \Gamma \, .
\end{align}
Consider the $\gamma \to \infty$ limit. In this limit $\tilde{W}_{x,\mu}$ reduces to the unit matrix: $\tilde{W}_{x,\mu}  \to \bm{1}$. 
Even after the global transformation $\Gamma \in \widetilde{\mathrm{SU(2)}}_\mathrm{global}$ this configuration is preserved: $\tilde{W}_{x,\mu}^{\prime}  \to \bm{1}$, because $\tilde{W}_{x,\mu}^{\prime} = \Gamma^\dagger \tilde{W}_{x,\mu} \Gamma \to \Gamma^\dagger \bm{1} \Gamma = \Gamma^\dagger \Gamma = \bm{1}$.
This means that the spontaneous breaking of $\widetilde{\mathrm{SU(2)}}_\mathrm{global}$ cannot be seen by using $\tilde{W}_{x,\mu}$. 
However, this does not mean that the spontaneous breaking of $\widetilde{\mathrm{SU(2)}}_\mathrm{global}$ does not occur.
We can find an appropriate operator which plays the role of the order parameter. Indeed, we can adopt $\bar{\tilde{\bm{n}}}$ constructed from the new color-direction field $\tilde{\bm{n}}_x$ for this purpose.  If a specific configuration $\bar{\tilde{\bm{n}}}_*$ of $\bar{\tilde{\bm{n}}}$ is preserved, it must satisfy $\bar{\tilde{\bm{n}}}_*^\prime :=\Gamma^\dagger \bar{\tilde{\bm{n}}}_* \Gamma =\bar{\tilde{\bm{n}}}_*$ which is equivalent to $\bar{\tilde{\bm{n}}}_* \Gamma =\Gamma \bar{\tilde{\bm{n}}}_* \Leftrightarrow [\bar{\tilde{\bm{n}}}_* , \Gamma]=0$, namely, $\bar{\tilde{\bm{n}}}_*$ must commute with any $\Gamma \in \widetilde{\mathrm{SU(2)}}_\mathrm{global}$.
Therefore, $\bar{\tilde{\bm{n}}}_*$  must be proportional to the unit matrix. However, this is impossible because $\bar{\tilde{\bm{n}}}_*$ is a Lie-algebra valued and traceless, and hence cannot be proportional to the unit matrix although $\bar{\tilde{\bm{n}}}_*$ is a two by two matrix, in contrast to $\tilde{W}_{x,\mu}$ which is a group-valued matrix. Thus, $\bar{\tilde{\bm{n}}}$ can be an order parameter to see the spontaneous breaking of the global $\widetilde{\mathrm{SU(2)}}_\mathrm{global}$ symmetry.
In the limit $\gamma \to \infty$ the reduction functional reduces to
\begin{align}
\tilde{F}_\mathrm{red} [\tilde{\bm{n}} ; \tilde{W}]
\to \sum_{x,\mu} \tr \left( \bm{1} - \tilde{\bm{n}}_x \tilde{\bm{n}}_{x+\mu} \right) ,
\label{red3}
\end{align}
and a constant color-direction field $\tilde{\bm{n}}_*$ becomes the solution of the reduction condition. 

Therefore, $\bar{\tilde{\bm{n}}}_*$ resulting from the constant configuration $\tilde{\bm{n}}_*$  indeed breaks the $\widetilde{\mathrm{SU(2)}}_\mathrm{global}$ symmetry in the limit $\gamma \to \infty$. 
Thus, $\bar{\tilde{\bm{n}}}$ can be used to construct the order parameter for the spontaneous breaking of $\widetilde{\mathrm{SU(2)}}_\mathrm{global}$. 

The OSFS results cannot be applied to the ``nonlocal'' operator with the support including infinite number of links or sites over the whole  lattice.  

It is important to remark that the spontaneous breaking of the global continuous symmetry can be searched by measuring the nonlocal operator with the support over all the sites on the lattice in the sense explained in the above. 
In other words, the new transition line does not represent a simple thermodynamics transition.

This operator was measured by obtaining the eigenvalues of the operator obtained by the diagonalization, which is achieved by applying the specific global rotation $\Gamma_*$.
This specific choice of the global rotation realizing the diagonalization of the nonlocal operator corresponds to breaking the global symmetry spontaneously. 
Thus the new formulation guarantees that the resulting spontaneous breaking of the global symmetry is gauge-invariant (gauge-independent) phenomenon, which also justifies the result obtained by taking the unitary gauge.


\section{EVALUATION OF $\langle \left\| \bm{R}_0 \right\|_1 \rangle$ AND $\langle \left\| \bm{R}_0 \right\|_2 \rangle$}

Let $\bm{r}_0$ be the random variable on the surface $S^2$ parametrized by the two angles $\theta, \phi$: 
\begin{align}
 \bm{r}_0 = (r_0^1, r_0^2, r_0^3) = (\sin\theta \cos\phi, \sin\theta \sin\phi, \cos\phi) \, .
\end{align}
The two angles $\theta, \phi$ follow the uniform distributions $\theta \sim U(0,\pi), \ \phi \sim U(0,2\pi)$.

Then we can calculate the expected values $E[r_0^A]$ and variances $V[r_0^A] = E[(r_0^A)^2] - E[r_0^A]^2$ for $A=1,2,3$:
\begin{align}
 E[r_0^A]
  &= \frac{1}{4\pi} \int_0^{2\pi} d\phi \int_{-1}^1 d(\cos\theta) r_0^A (\theta,\phi) = 0 \, , \\
 V[r_0^A]
  &= \frac{1}{4\pi} \int_0^{2\pi} d\phi \int_{-1}^1 d(\cos\theta) {r_0^A (\theta,\phi)}^2 = \frac{1}{3} \, ,
\end{align}
and $E[R_0^A]$ and $V[R_0^A]$ for $\bm{R}_0 = \frac{1}{V} \sum_x \bm{r}_{0,x}$:
\begin{align}
 E[R_0^A]
  &= \frac{1}{V} \sum_x E[r_0^A] = 0 \, , \notag\\
 V[R_0^A]
  &= \frac{1}{V^2} \sum_x V[r_0^A] = \frac{1}{3V} \, .
\end{align}

According to the standard argument, it is shown that $R_0^A$ follows a normal distribution $\mathcal{N} \left( \mu, \sigma^2 \right)$: $R_0^A \sim \mathcal{N} \left( \mu = 0 , \sigma^2 = \frac{1}{3V} \right)$ due to the central limit theorem.

We represent the components as $(x,y,z) := \bm{R}_0 = (R_0^1, R_0^2, R_0^3)$, and the Gaussian distribution as
\begin{align}
 g(x) = \frac{1}{\sqrt{2\pi \sigma^2}} \exp \left( -\frac{x^2}{2\sigma^2} \right)
\end{align}
in the subsequent evaluations. $g(y), g(z)$ are also defined in a similar way.\\

\noindent
(i) Evaluation of $\langle \left\| \bm{R}_0 \right\|_1 \rangle$

For the 1-norm $\left\| \bm{R}_0 \right\|_1 = |x|+|y|+|z|$, the expected value $E[\left\| \bm{R}_0 \right\|_1]$ and variance $V[\left\| \bm{R}_0 \right\|_1]$ can be evaluated as
\begin{align}
 &E[\left\| \bm{R}_0 \right\|_1]
  = \int dxdydz \left( |x|+|y|+|z| \right) g(x)g(y)g(z) \notag\\
  &= 3\int_{-\infty}^{\infty} dx |x| g(x) \notag\\
  &= 3\sqrt{\frac{2}{\pi}} \sigma
    = \sqrt{\frac{6}{\pi}} \frac{1}{\sqrt{V}}
    \equiv \mu_1 \, ,
\end{align}
\begin{align}
 &V[\left\| \bm{R}_0 \right\|_1]
  = E[\left\| \bm{R}_0 \right\|_1^2] - \mu_1^2 \notag\\
  &= \int dxdydz \left( |x|+|y|+|z| \right)^2 g(x)g(y)g(z) - \mu_1^2 \notag\\
  &= 3\int_{-\infty}^{\infty} dx x^2 g(x) + 6\left( \int_{-\infty}^{\infty} dx |x| g(x) \right)^2 - \mu_1^2 \notag\\
  &= 3 \cdot \sigma^2 + 6\left( \sqrt{\frac{2}{\pi}} \sigma \right)^2 - \left( 3\sqrt{\frac{2}{\pi}} \sigma \right)^2 \notag\\
  &= 3\left( 1-\frac{2}{\pi} \right) \sigma^2
    = \left( 1-\frac{2}{\pi} \right) \frac{1}{V}
    \equiv \sigma_1^2 \, .
\end{align}

For the sampling average $\langle \left\| \bm{R}_0 \right\|_1 \rangle = \frac{1}{N} \sum_{i=1}^N \left\| \bm{R}_0 \right\|_{1,i}$ based on $N$ samples, the expected value $E[\langle \left\| \bm{R}_0 \right\|_1 \rangle]$ and mean squared error $\delta^2 [\langle \left\| \bm{R}_0 \right\|_1 \rangle]$ are evaluated as
\begin{align}
 E[\langle \left\| \bm{R}_0 \right\|_1 \rangle]
  &= \frac{1}{N} \sum_{i=1}^N E[\left\| \bm{R}_0 \right\|_{1,i}]
  = \mu_1 \notag\\
  &= \sqrt{\frac{6}{\pi}} \frac{1}{\sqrt{V}} \, , \\
 \delta^2 [\langle \left\| \bm{R}_0 \right\|_1 \rangle]
  &= E[\left( \langle \left\| \bm{R}_0 \right\|_1 \rangle - \mu_1 \right)^2]
  = \left( \frac{\sigma_1}{\sqrt{N}} \right)^2 \notag\\
  &= \left( \sqrt{1-\frac{2}{\pi}} \frac{1}{\sqrt{V}} \frac{1}{\sqrt{N}} \right)^2 \, .
\end{align}

\noindent
(ii) Evaluation of $\langle \left\| \bm{R}_0 \right\|_2 \rangle$

For the 2-norm $\left\| \bm{R}_0 \right\|_2 = \left( x^2+y^2+z^2 \right)^{1/2}$, the expected value $E[\left\| \bm{R}_0 \right\|_2]$ and variance $V[\left\| \bm{R}_0 \right\|_2]$ can be evaluated as
\begin{align}
 &E[\left\| \bm{R}_0 \right\|_2]
  = \int dxdydz \left( x^2+y^2+z^2 \right)^{1/2} g(x)g(y)g(z) \notag\\
  &= \frac{1}{(2\pi \sigma^2)^{3/2}} \int_0^{2\pi} d\varphi \int_{-1}^1 d(\cos\theta) \int_0^{\infty} dr r^3 e^{-r^2/2\sigma^2} \notag\\
  &= \sqrt{\frac{8}{\pi}} \sigma
    = \sqrt{\frac{8}{3\pi}} \frac{1}{\sqrt{V}}
    \equiv \mu_2 \, ,
\end{align}
\begin{align}
 &V[\left\| \bm{R}_0 \right\|_2]
  = E[\left\| \bm{R}_0 \right\|_2^2] - \mu_2^2 \notag\\
  &= \int dxdydz \left( x^2+y^2+z^2 \right) g(x)g(y)g(z) - \mu_2^2 \notag\\
  &= 3\int_{-\infty}^{\infty} dx x^2 g(x) - \mu_2^2 \notag\\
  &= 3\left( 1-\frac{8}{3\pi} \right) \sigma^2
    = \left( 1-\frac{8}{3\pi} \right) \frac{1}{V}
    \equiv \sigma_2^2 \, ,
\end{align}

For the sampling average $\langle \left\| \bm{R}_0 \right\|_2 \rangle = \frac{1}{N} \sum_{i=1}^N \left\| \bm{R}_0 \right\|_{2,i}$ based on $N$ samples, the expected value $E[\langle \left\| \bm{R}_0 \right\|_2 \rangle]$ and mean squared error $\delta^2 [\langle \left\| \bm{R}_0 \right\|_2 \rangle]$ are evaluated as
\begin{align}
 E[\langle \left\| \bm{R}_0 \right\|_2 \rangle]
  &= \frac{1}{N} \sum_{i=1}^N E[\left\| \bm{R}_0 \right\|_{2,i}]
  = \mu_2 \notag\\
  &= \sqrt{\frac{8}{3\pi}} \frac{1}{\sqrt{V}} \, , \\
 \delta^2 [\langle \left\| \bm{R}_0 \right\|_2 \rangle]
  &= E[\left( \langle \left\| \bm{R}_0 \right\|_2 \rangle - \mu_2 \right)^2]
  = \left( \frac{\sigma_2}{\sqrt{N}} \right)^2 \notag\\
  &= \left( \sqrt{1 - \frac{8}{3\pi}} \frac{1}{\sqrt{V}} \frac{1}{\sqrt{N}} \right)^2 \, .
\end{align}

It should be noticed that the above expected values vanish in the limit $V \to \infty$:
\begin{align}
 \lim_{V \to \infty} E[\langle \left\| \bm{R}_0 \right\|_1 \rangle]
  = 0 \, , \quad
 \lim_{V \to \infty} E[\langle \left\| \bm{R}_0 \right\|_2 \rangle]
  = 0 \, .
\end{align}

Moreover, we can give the evaluation for the case $V=8^4, N=100$ as
\begin{align}
 \langle \left\| \bm{R}_0 \right\|_1 \rangle &= (2.16 \pm 0.09) \times 10^{-2} \, , \notag\\
 \langle \left\| \bm{R}_0 \right\|_2 \rangle &= (1.44 \pm 0.06) \times 10^{-2} \, ,
\end{align}
and for the case $V=16^4, N=100$ as
\begin{align}
 \langle \left\| \bm{R}_0 \right\|_1 \rangle &= (5.40 \pm 0.24) \times 10^{-3} \, , \notag\\ 
 \langle \left\| \bm{R}_0 \right\|_2 \rangle &= (3.60 \pm 0.15) \times 10^{-3} \, .
\end{align}

\section*{Acknowledgments}
This work was supported by JST, the establishment of university fellowships toward the creation of science technology innovation, Grant No. JPMJFS2107.
This work was also supported by Grant-in-Aid for Scientific Research, JSPS KAKENHI Grant No. (C) No.23K03406. 
This research was supported in part by the Multidisciplinary
Cooperative Research Program in CCS, University of Tsukuba. 
The numerical simulation is supported by the High Energy Accelerator Research Organization (KEK).

\end{document}